%% file: agent-first-canonical-code.tex
\documentclass[conference]{IEEEtran}

\usepackage{cite}
\usepackage{amsmath}
\usepackage{microtype}
\usepackage{booktabs}
\usepackage{array}
\usepackage{tabularx}
\usepackage{makecell}
\usepackage{siunitx}
\usepackage{url}
\usepackage{xurl}
\usepackage{graphicx}
\usepackage[table]{xcolor}
\usepackage{tikz}
\usetikzlibrary{arrows.meta,positioning,shapes.geometric}
\usepackage[hidelinks]{hyperref}
\hypersetup{
  pdftitle={No Accidental Software: Agent-First Canonical Code for Human-Code Entropy Reduction and 30x--500x Lower Frontier-Model Requirements},
  pdfauthor={Jepson Taylor},
  pdfsubject={Agent-first canonical code training, corpus compression, behavior cells, proof-carrying change objects, reasoning compression, and verified correct-change economics},
  pdfkeywords={canonical code, code language models, agent-first programming, corpus compression, software training data, behavior cells, proof-carrying change objects}
}

\newcommand{\canonprofile}{versioned limited-stack product profile}

\newcommand{\mfdl}{\textsc{mfdl}}
\newcommand{\measured}{\textsc{Measured}}
\newcommand{\targetclaim}{\textsc{Near-term target}}
\newcommand{\centralclaim}{\textsc{Central hypothesis}}
\newcommand{\moonshotclaim}{\textsc{Moonshot}}
\definecolor{nhConservative}{RGB}{20,125,72}
\definecolor{nhCentral}{RGB}{52,72,94}
\definecolor{nhAggressive}{RGB}{190,44,44}
\definecolor{nhBlue}{RGB}{18,90,166}
\definecolor{nhPurple}{RGB}{108,55,165}
\definecolor{nhGold}{RGB}{156,101,0}
\definecolor{nhPaleBlue}{RGB}{235,244,255}
\definecolor{nhPaleGreen}{RGB}{235,250,241}
\definecolor{nhPaleRed}{RGB}{255,239,239}
\definecolor{nhPaleGold}{RGB}{255,248,230}
\definecolor{nhPalePurple}{RGB}{245,240,255}
\newcommand{\conrange}[1]{{\color{nhConservative}#1}}

\newcommand{\aggrange}[1]{{\color{nhAggressive}#1}}
\newcommand{\conhead}{\conrange{\textbf{Conservative}}}
\newcommand{\agghead}{\aggrange{\textbf{Aggressive}}}
\newcommand{\bigclaim}[2]{\textcolor{#1}{\textbf{#2}}}

\newcommand{\oracle}{\mathcal{O}}
\newcommand{\canonmap}{\Phi}

\title{\textbf{No Accidental Software}\\
Agent-First Canonical Code for Human-Code Entropy Reduction and 30$\times$--500$\times$ Lower Frontier-Model Requirements}

\author{
\IEEEauthorblockN{Jepson Taylor}
\IEEEauthorblockA{NeverHuman Research}
}

\begin{document}
\twocolumn[
\begin{@twocolumnfalse}
\maketitle

\begin{abstract}
\textbf{The number is the hook, but the denominator is the science: frontier coding models are paying to learn human-code entropy, not just code.} Raw human repositories contain behavior, incidents, tests, edge cases, migrations, product judgment, and operational scar tissue worth preserving, but those signals are entangled with accidental representation: language fashion, framework churn, naming drift, folder folklore, duplicated contracts, generated-source confusion, dependency rituals, continuous-integration dialects, weak proof routes, and review customs built for humans rather than agents. A coding agent pays for this entropy four times: during training, during context gathering, during reasoning/tool/retry loops, and during human review. This paper proposes \textbf{agent-first canonical code}: a governed, proof-carrying substrate that ports routine product software into canonical profiles, behavior cells, generated truth, typed change algebra, proof lanes, constrained edit grammars, reasoning digests, semantic patch cells, runtime negative memory, and proof-carrying change objects before training and before agent operation.

The claim is deliberately ambitious, but every range has a denominator. Detailed conservative, central, and aggressive bands are hypotheses reported in the tables, not measured frontier results, and they must not be multiplied into a miracle. The decisive endpoint is all-in cost per verified correct change: source tokens, context, reasoning, tools, verification, security, provenance, review, failed loops, downstream defects, and amortized foundry cost under the same acceptance oracle.

The theoretical endpoint is the No-Accident Horizon: under a declared oracle and supported routine-product distribution, all removable accident can vanish only until the residual floor of novelty, evidence, governance, risk, and future optionality remains; a mature foundry therefore has a strongest defensible planning limit near 100$\times$ all-in verified-change cost reduction, not an unrestricted all-code guarantee.

The theoretical spine is quotienting software by behavior equivalence under a declared oracle. Many behavior-equivalent human encodings should collapse to one governed representative plus evidence, while residual novelty, legal risk, security exposure, and domain mismatch remain explicit in a disposition ledger. The limit is \textsc{Minimum Functional Description Length}: the shortest canonical behavior specification, evidence bundle, proof obligations, and renderer that produce a working system under behavior, security, migration, provenance, and review constraints. Preliminary QLoRA evidence on Qwen2.5-Coder-14B shows that 64{,}088 canonically translated trajectories are learnable and suppress tested forbidden-language markers; it does \emph{not} establish behavior preservation, scaling economics, or verified-change cost. The contribution is a falsifiable research program whose decisive endpoint is amortized cost per verified correct change.
\end{abstract}


\noindent\fcolorbox{nhGold}{nhPaleGold}{\parbox{0.965\textwidth}{\textbf{One-sentence thesis.} Raw repositories force models to learn both product behavior and accidental implementation orbits; canonical proof-carrying behavior lets them spend training, inference, review, and verification budget on verified change instead of rediscovering local software folklore.}}

\vspace{0.5em}
\noindent\fcolorbox{nhPurple}{nhPalePurple}{\parbox{0.965\textwidth}{\textbf{What would make this undeniable.} On paired raw versus canonical repositories from the same lineage, the same model must spend materially fewer source, context, reasoning, tool, retry, and reviewer tokens to produce an accepted proof-carrying change. Only after that same-model result holds should specialist scaling curves be allowed to justify the 30$\times$--150$\times$ central training-token claim and the 8.3$\times$--10$\times$ dense-active serving-speed claim.}}

\vspace{0.75em}
\end{@twocolumnfalse}
]

\begin{table*}[!t]
\caption{The Hook: Conservative-to-Aggressive Payoff Ranges by Measurement Denominator}
\label{tab:hook-payoff}
\centering
\footnotesize
\renewcommand{\arraystretch}{1.16}
\begin{tabularx}{\textwidth}{@{}p{0.21\textwidth}>{\columncolor{nhPaleGreen}}c>{\columncolor{nhPaleBlue}}c>{\columncolor{nhPaleRed}}cX@{}}
\toprule
\textbf{Outcome denominator} & \conhead & \textbf{Central} & \agghead & \textbf{What must be measured} \\
\midrule
Training tokens to fixed accepted-change target & \bigclaim{nhConservative}{$10\times$--$30\times$} & \bigclaim{nhBlue}{$30\times$--$150\times$} & \bigclaim{nhAggressive}{$150\times$--$1{,}000\times$} & Paired raw/canonical scaling curves to the same hidden-test accepted-change capability. \\
Context, reasoning, tool, and retry tokens per verified change & \bigclaim{nhConservative}{$3\times$--$10\times$} & \bigclaim{nhBlue}{$10\times$--$100\times$} & \bigclaim{nhAggressive}{$100\times$--$10{,}000\times$} & Files opened, planning tokens, hidden reasoning budget where measurable, tool calls, failed repair loops, and validation reruns. \\
Cost per verified correct change & \bigclaim{nhConservative}{$3\times$--$10\times$} & \bigclaim{nhBlue}{$10\times$--$50\times$} & \bigclaim{nhAggressive}{$50\times$--$1{,}000\times$} & End-to-end dollars including source/context/action tokens, serving, verification, security/provenance, review, downstream defects, and amortized foundry cost. \\
Same-model verified-change wall-clock & \bigclaim{nhConservative}{$2\times$--$5\times$} & \bigclaim{nhBlue}{$5\times$--$20\times$} & \bigclaim{nhAggressive}{$20\times$--$100\times$} & End-to-end elapsed time from issue receipt to accepted proof-carrying patch under equal tool and reviewer budgets. \\
Dense-active inference speed scenario & \bigclaim{nhConservative}{$3\times$--$5\times$} & \bigclaim{nhBlue}{$8.3\times$--$10\times$} & \bigclaim{nhAggressive}{$10\times$--$20\times$} & Conditional only: matched-quality canonical specialist versus 1T-class dense-active baseline on supported canonical work. \\
Dense-active infrastructure cost/token & \bigclaim{nhConservative}{67\%--80\% lower} & \bigclaim{nhBlue}{88\%--90\% lower} & \bigclaim{nhAggressive}{90\%--95\% lower} & Active-parameter, memory, batching, utilization, MoE, context, and verification-overhead accounting. \\
Effective action/representation space & \bigclaim{nhConservative}{$10\times$--$40\times$} & \bigclaim{nhBlue}{$40\times$--$150\times$} & \bigclaim{nhAggressive}{$100\times$--$300\times$} & Entropy of valid behavior-equivalent encodings and legal edit choices after contracts, profiles, and proof lanes are fixed. \\
Foundry amortization and reuse & \bigclaim{nhConservative}{$2\times$--$5\times$} & \bigclaim{nhBlue}{$10\times$--$100\times$} & \bigclaim{nhAggressive}{$100\times$--$10{,}000\times$} & Reuse of behavior cells, semantic patches, negative memory, generated projections, and proof receipts across independent product lineages. \\
\bottomrule
\end{tabularx}
\end{table*}

\begin{table*}[!t]
\caption{External Facts the Canonical-Code Thesis Must Explain}
\label{tab:external-facts}
\centering
\footnotesize
\renewcommand{\arraystretch}{1.14}
\begin{tabularx}{\textwidth}{p{0.21\textwidth}X X}
\toprule
\textbf{Observed fact} & \textbf{Why it matters} & \textbf{Canonical-code interpretation} \\
\midrule
Public software is massive and multi-language: The Stack v2 contains over 3B files in 600+ languages and StarCoder2 trains on 3.3T--4.3T tokens across 619 languages~\cite{theStackV2,starcoder2}. & Broad code models pay to learn the whole human archive. & The archive is behavior raw material, not necessarily the optimal training substrate. \\
Duplication is structural: D\'ej\`aVu found 85M unique files among 428M files across 4.5M non-fork projects, and Jupyter studies found more than 70\% exact snippet copies~\cite{dejavu2017,jupyterClones2020}. & Human code repeats behavior and representation. & Cells and semantic patches should capture repeated behavior directly. \\
Agent benchmarks are brittle: SWE-bench Verified improves curation, but UTBoost found insufficient tests and erroneous pass labels in SWE-bench-family results~\cite{swebenchVerified,utboost}. & Passing tests is not equivalent to preserving behavior. & Ports and changes need evidence tiers, hidden tests, fuzzing, replay, mutation tests, and proof receipts. \\
Agent cost matters: resource-constrained evaluation shows why resolve rate alone hides token, time, and failure cost~\cite{metr2025,metr2026}. & Accuracy without cost accounting can be economically misleading. & The denominator must be cost per verified correct change. \\
CI/security sprawl is real: recent GitHub Actions studies report workflow heterogeneity and security-practice gaps, and GitGuardian reported 28.65M new public-GitHub secrets in 2025~\cite{ghaHeterogeneity2025,ghaEvolution2026,ghaSecurityNDSS2026,gitguardian2026}. & Repository operations are part of the software behavior surface. & CI, secrets, permissions, and release policy must become canonical substrate objects. \\
\bottomrule
\end{tabularx}
\end{table*}

\begin{table*}[!t]
\caption{Front-Door Measurement Ledger}
\label{tab:frontledger}
\centering
\footnotesize
\renewcommand{\arraystretch}{1.14}
\begin{tabularx}{\textwidth}{p{0.18\textwidth}p{0.22\textwidth}X X}
\toprule
\textbf{Question} & \textbf{Primary denominator} & \textbf{First test} & \textbf{Failure mode} \\
\midrule
Does the substrate help before new training? & Same-model cost per accepted change & Broad model on paired raw vs. canonical repositories with the same issue lineage, hidden tests, and reviewer rubric. & Canonical form looks cleaner but does not reduce files opened, context/reasoning tokens, tool calls, failed lanes, or review. \\
Does canonical data train more efficiently? & Tokens-to-target accepted-change capability & Paired scaling curves from the same raw/canonical lineage. & Canonical data requires similar tokens, similar model scale, or larger models for the same accepted-change target. \\
Does the serving denominator improve? & Dense-active infrastructure cost/token and token/sec at matched quality & Compare a broad dense-active baseline against a canonical specialist on supported canonical work. & Throughput gains disappear after batching, memory, context, verification, or quality corrections. \\
Does behavior survive porting? & Preservation tier and hidden oracle pass rate & Differential tests, replay, fuzzing, contracts, migration checks, security checks, and human review. & Important compatibility ghosts are lost or security/migration regressions rise. \\
Does the economics close? & All-in cost per verified correct change & Foundry amortization plus training, serving, tools, verification, failed loops, and review. & Foundry and verification overhead consume the gains. \\
\bottomrule
\end{tabularx}
\end{table*}


\input{sections/01_introduction.tex}
\input{sections/01a_positioning_and_theoretical_move.tex}
\input{sections/02_definitions.tex}
\input{sections/02a_correct_change_information_theory.tex}
\input{sections/03_human_code_entropy.tex}
\input{sections/04_assumption_a_profile.tex}
\input{sections/05_assumption_b_foundry.tex}
\input{sections/06_assumption_c_behavior_preservation.tex}
\input{sections/07_assumption_d_behavior_cells.tex}
\input{sections/08_empirical_validation.tex}
\input{sections/09_assumption_e_compression_theory.tex}
\input{sections/09a_behavior_ir_typed_change_algebra.tex}
\input{sections/10_assumption_i_reasoning_compression.tex}
\input{sections/10_assumption_f_economics.tex}
\input{sections/11_assumption_g_role_moe.tex}
\input{sections/12_assumption_h_evaluation.tex}
\input{sections/12a_cost_ledger_field_trial.tex}
\input{sections/13_risks_and_threat_model.tex}
\input{sections/13_compression_frontier.tex}
\input{sections/14_future_vision.tex}
\input{sections/14a_first_six_experiments.tex}
\input{sections/14b_frontier_extensions.tex}
\input{sections/14c_no_accident_horizon.tex}
\input{sections/14_limits_and_conclusion.tex}

\appendices
\input{compression_catalog.tex}

\bibliographystyle{IEEEtran}
\bibliography{references}

\end{document}

%% file: sections/01_introduction.tex
\section{Introduction}

The conservative claim is already large: 10$\times$--30$\times$ fewer training tokens to a fixed accepted-change capability, 3$\times$--10$\times$ fewer context/reasoning/tool/retry tokens per verified change, and 3$\times$--10$\times$ lower all-in cost per verified correct change after foundry amortization. The central claim is larger: 30$\times$--150$\times$ training-token efficiency, 10$\times$--100$\times$ fewer reasoning/action/retry tokens, and 10$\times$--50$\times$ lower verified-change cost on supported canonical product work. The denominator comes first because the thesis is not ``better style.'' It is that frontier coding models are paying to learn the wrong object.

Frontier coding models are trained on a software record built by humans for humans. That record contains valuable behavior, but it is not a canonical representation of software. It is a survival archive: hundreds of languages, millions of local conventions, duplicated frameworks, inconsistent naming, weak tests, dependency sprawl, build folklore, continuous-integration (CI) dialects, repository settings, access policies, and release rituals that were never designed for autonomous agents. A coding agent pays for that archive four times: during training, during context gathering, during reasoning/tool/retry loops, and during human review.

The end-of-road story is stark. In the conservative version, canonicalization is an engineering discipline that makes coding agents cheaper and more reliable by cutting routine repository search and failed repair loops. In the central version, it becomes a new data substrate: fewer legal encodings, fewer legal edit paths, and lower tokens-to-target capability. In the aggressive version, routine product software approaches a behavior-genome regime in which most CRUD, auth, policy, migration, deployment, observability, billing, notification, workflow, and integration behavior is assembled from proof-carrying cells, while bespoke code is reserved for minimum viable novelty. The paper is ambitious because the upside is enormous; it is defensible only because every large number is tied to a named denominator and a falsification test.

The scale is enormous. GitHub reported 395 million public and open-source repositories in Octoverse 2025, roughly 230 new repositories per minute, and a major shift in the most-used language~\cite{octoverse2025}. The Stack v2 reports 67.5TB full, 32.1TB deduplicated, roughly 900B train-full tokens, over 3B files, and 658 languages~\cite{theStackV2}. StarCoder2 was trained on 3.3T--4.3T tokens across 619 programming languages~\cite{starcoder2}. Kimi K2 reports a 1T-total-parameter mixture-of-experts model with 32B active parameters and 15.5T pretraining tokens~\cite{kimiK2,kimiDocs}. Modern code-model work shows the same field-level direction: more languages, longer context, more tokens, more data curation, and more agentic training rather than a smaller human-code target~\cite{deepseekCoder,qwenCoder}.

The redundancy is also enormous. D\'{e}j\`{a}Vu analyzed 4.5 million non-fork GitHub projects and found that only 85 million of 428 million files were unique---about 70\% were clones of previously created files~\cite{dejavu2017}. Jupyter notebook code shows similar patterns: more than 70\% of code snippets were exact copies~\cite{jupyterClones2020}. Software has long been known to be highly repetitive and predictable~\cite{hindleNaturalness}.

The thesis is stronger than filtering or deduplication: the available training code should be \emph{ported} into an agent-first canonical standard before training. Filtering selects cleaner examples from the old distribution. Canonical porting changes the distribution and the action substrate. The standard constrains language roles, file grammar, naming, dependency ownership, generated boundaries, database truth, continuous integration, repository policy, build/test commands, migration policy, proof lanes, and agent-readable edit scope. Observable behavior is preserved, rejected, or tiered according to an explicit evidence envelope; arbitrary human degrees of freedom collapse only when they are not part of that envelope. The external-reader claim is precise: human repositories are improvable because they mix durable behavior with accidental representation, not because human behavior, product judgment, operational lessons, or edge cases are disposable.

This is a deliberately strong position: for every supported product-software concern there should be one governed way to express it, and for other software classes there should be a secondary governed profile rather than local preference. Humans do not merely choose different surface styles; they create mutually incompatible local dialects for routing, testing, continuous integration, authorization, secrets, migrations, feature flags, deployment, observability, and review. Those dialects may work locally, but they are training noise for agents and reasoning overhead at inference time. For supported classes, the falsifiable claim is that nearly every nontrivial human-authored repository contains reducible accidental degrees of freedom: duplicate behavior, noncanonical representation, generated-truth drift, unconstrained edit surface, missing proof receipts, or ungoverned policy.

The decisive experiment is not a better-looking repository. It is a same-model, same-lineage comparison: original human repository versus canonical repository, same issue, same behavior contract, same hidden tests, same reviewer rubric, and a complete ledger of files opened, context tokens, reasoning tokens, tool calls, invalid edits, failed proof lanes, review burden, wall time, and accepted-patch rate. If a broad model performs materially better on the canonical repository before a canonical specialist is trained, then substrate compression is real. If it does not, the thesis fails early.

This paper makes the case through nine explicit assumptions, each with falsification criteria, a compression frontier catalog, preliminary training evidence, and a paired evaluation design. The core falsifiable thesis is simple: after controlling for behavior and task, canonical proof-carrying software should lower same-model search cost now, lower tokens-to-target accepted-change capability under paired scaling curves later, and lower amortized cost per verified correct change in production field trials. If those three gates fail, no title, figure, or compression catalog rescues the claim.


%% file: sections/01a_positioning_and_theoretical_move.tex
\section{Positioning: What Existing Work Proves, and What It Does Not}

Adjacent work strongly supports the premise that software contains compressible regularity, but it does not yet prove the thesis in this paper. Naturalness results show that code is unusually repetitive and predictable~\cite{hindleNaturalness}; clone studies show massive exact and near-exact duplication in public repositories~\cite{dejavu2017,jupyterClones2020}; modern code LLMs show that trillion-token corpora can learn broad programming competence~\cite{starcoder2,qwenCoder,deepseekCoder,kimiK2}; and repository-level benchmarks show that agents can now operate over real projects, although contamination, weak tests, and distribution mismatch make static benchmarks fragile~\cite{swebench,swebenchVerified,sweRebench,utboost}. This paper's stronger claim is different: the target distribution itself is wrong. The right move is not only higher-quality filtering, more synthetic tasks, or larger context windows. It is to transform the code archive into a canonical behavior-and-change substrate before training.

The strongest allies are not only code LLM papers. They are program-analysis and programming-language systems that already separate meaning from textual accident. OpenAPI and Protocol Buffers show that one source of truth can generate multiple projections of an interface~\cite{openapi31,protobufOverview}. MLIR shows that domain-specific dialects can lower through reusable compiler infrastructure rather than forcing every domain into one flat representation~\cite{lattnerMLIR}. Coccinelle shows that semantic patches can encode collateral evolutions more compactly than hand-written diffs~\cite{coccinelle}. CodeQL shows that code can become queryable semantic data for large-scale vulnerability reasoning~\cite{codeql}. E-graphs show how equivalence classes of programs can be represented compactly for rewrite search~\cite{eggEgraphs}. DreamCoder and library-learning systems show that repeated programs can be compressed into reusable abstractions that improve search~\cite{dreamcoder}. Agent-first canonical code combines these threads into one substrate-level bet: learn from human software history, but do not force future agents to imitate its incidental coordinates.

This positioning also defines the paper's burden of proof. A reviewer should not accept the claim because the standard is elegant. The claim earns belief only if paired experiments show lower cost per verified correct change, lower tokens-to-target accepted-change capability, lower action branching, preserved behavior at declared evidence tiers, and lower infrastructure cost/token where model-size claims are invoked. Conversely, if canonicalization removes useful implementation diversity, harms robustness, fails to amortize foundry cost, or merely moves entropy into metadata, the thesis should be rejected or narrowed.

\section{The Missing Theoretical Move: Train on the Quotient, Not the Orbit}
\label{sec:quotient}

The deepest version of the thesis is not that code can be standardized. It is that software learning should be quotiented by behavior. Let $P$ be the set of source-level programs, repositories, configurations, schemas, tests, and deployment artifacts. Let $O$ be a declared behavior oracle: public tests, hidden tests, trace equivalence, API contracts, migration replay, security policies, performance envelopes, accessibility checks, runtime invariants, and accepted incompatibility dispositions. Define an equivalence relation
\begin{equation}
p_1 \sim_O p_2 \quad \Longleftrightarrow \quad O(p_1)=O(p_2)
\end{equation}
up to the declared preservation tier. The human software corpus samples many points in the syntactic orbit $[p]_O$: Python/FastAPI, Java/Spring, Node/Express, Ruby/Rails, Go/Gin, C\#/ASP.NET, hand-written DTOs, generated DTOs, imperative migrations, declarative migrations, bespoke CI, and ad hoc review rules can all encode the same product behavior. A raw code model learns both the quotient $P/{\sim_O}$ and the orbit noise inside each class. A canonical foundry instead learns a versioned normal-form map
\begin{equation}
\kappa: [p]_O \rightarrow (s,e,r,d),
\end{equation}
where $s$ is the shortest accepted canonical specification, $e$ is the evidence bundle, $r$ is the renderer/generator, and $d$ is the disposition ledger for non-preserved behavior, legal risk, security defects, or domain mismatch.

This distinction clarifies why the proposal is more ambitious than deduplication. Deduplication removes copied files. Near-deduplication improves pretraining quality by removing repeated text. Canonical quotienting removes degrees of freedom that should not be in the model's target language at all. The model is no longer asked to learn every valid implementation orbit; it is asked to learn canonical representatives plus the proof obligations that make the representative behaviorally acceptable.

\begin{table*}[t]
\caption{From Raw-Code Imitation to Behavior-Quotient Learning}
\label{tab:quotient}
\centering
\footnotesize
\begin{tabularx}{\textwidth}{p{0.18\textwidth}X X}
\toprule
Layer & Raw-code learning target & Canonical quotient target \\
\midrule
Representation & All languages, frameworks, layouts, generated artifacts, dependency rituals, and local repair customs. & One governed representative per behavior class, plus explicit profile escape hatches. \\
Generalization & Infer semantics through many surface encodings. & Learn behavior primitives, typed deltas, proof obligations, and legal edit forms. \\
Search & Explore file trees and patch locations in a high-entropy orbit. & Route to cells, owned files, generated truth, and proof lanes. \\
Verification & Interpret arbitrary tests, logs, CI conventions, and reviewer expectations. & Consume structured proof receipts and behavior-oracle deltas. \\
Memory & Remember past trajectories as chat/log traces. & Compile successful and failed trajectories into reasoning digests, semantic patch cells, and negative memory. \\
Economic unit & Patch accepted after ad hoc reasoning and review. & Proof-carrying change accepted under a cost ledger. \\
\bottomrule
\end{tabularx}
\end{table*}

The theory also explains why the proposed breakthrough can coexist with modern scaling laws. Scaling-law work implies that data quality and token efficiency matter; Chinchilla-style compute-optimal training links model size and training tokens under a compute budget~\cite{hoffmannChinchilla}. Canonical code does not repeal scaling laws. It changes the data distribution so that a token is closer to irreducible behavior and farther from accidental representation. If the canonical token stream has lower conditional entropy and fewer irrelevant action branches, the same physical training and inference budget should buy more verified software capability.

A useful analogy is compiler IR. LLVM and MLIR do not eliminate source languages; they provide structured intermediate forms that make analysis, lowering, optimization, and reuse tractable~\cite{lattnerLLVM,lattnerMLIR}. The canonical-code claim is that product software needs a higher IR: entities, permissions, policies, effects, state machines, migrations, API contracts, observability, proof obligations, rollout envelopes, and runtime invariants. Source code is then a projection, not the source of truth. Coccinelle's semantic patches show that many source changes are better represented as semantic transformations than text diffs~\cite{coccinelle}; e-graphs show that equivalence classes can be exploited computationally~\cite{eggEgraphs}. The missing leap is to make these ideas the training substrate for software agents.

\textbf{The falsifiable prediction:} after controlling for behavior and task, canonical quotient learning should reduce entropy, search, and proof cost before any new specialist model is trained. If a broad frontier model does not become cheaper and more reliable when the same software lineage is presented through the canonical representative, the quotient thesis is wrong.


%% file: sections/02_definitions.tex
\section{Definitions and Claim Denominators}

The paper separates public facts, preliminary measurements, near-term targets, central hypotheses, and moonshot targets.

Human-code entropy is the residual representation uncertainty left after behavior, contracts, and execution environment are fixed:
\begin{equation}
\begin{aligned}
H_{\mathrm{human}} ={}& H(\mathrm{representation} \mid \\
&\mathrm{behavior},\mathrm{contracts},\mathrm{environment}).
\end{aligned}
\end{equation}
A canonical substrate is useful only if it reduces this conditional entropy without erasing behavior evidence, safety constraints, provenance, compatibility ghosts, or deliberate diversity.

We report the following distinct reductions:
\begin{align}
R_{\text{source}} &= \frac{\text{raw source tokens}}{\text{canonical source tokens}}, \\
R_{\text{entropy}} &= \frac{H(\text{raw valid forms})}{H(\text{canonical valid forms})}, \\
R_{\text{action}} &= \frac{\text{raw attempted/legal actions}}{\text{canonical legal actions}}, \\
R_{\text{train}} &= \frac{\text{raw tokens to target accuracy}}{\text{canonical tokens to target accuracy}}, \\
R_{\text{train-time}} &= \frac{\text{raw wall-clock to target accuracy}}{\text{canonical wall-clock to target accuracy}}, \\
R_{\text{speed}} &= \frac{\text{raw wall-clock per verified change}}{\text{canonical wall-clock per verified change}}, \\
R_{\text{cost/token}} &= \frac{\text{raw dense-active infrastructure cost/token}}{\text{canonical dense-active infrastructure cost/token}}, \\
R_{\text{reason}} &= \frac{T^{\text{raw}}_{\text{reason+tool+plan+retry}}/\Delta}{T^{\text{canon}}_{\text{reason+tool+plan+retry}}/\Delta}, \\
R_{\text{retry}} &= \frac{L^{\text{raw}}_{\text{failed proof/repair}}/\Delta}{L^{\text{canon}}_{\text{failed proof/repair}}/\Delta}, \\
R_{\text{invalid}} &= \frac{A^{\text{raw}}_{\text{invalid edit}}/\Delta}{A^{\text{canon}}_{\text{invalid edit}}/\Delta}, \\
R_{\text{cost}} &= \frac{\text{raw cost per verified correct change}}{\text{canonical cost per verified correct change}}.
\end{align}
Here $\Delta$ denotes an accepted change. $R_{\text{cost}}$ is the all-in version of the earlier $R_{\text{change}}$ notation, and $R_{\text{cost/token}}$ is only the dense-active serving denominator. At fixed accelerator throughput, $R_{\text{train-time}} \approx R_{\text{train}}$. Effective raw-corpus-equivalent training throughput is $S_{\text{eff}} = S_{\text{physical}}R_{\text{train}}$, so a 1M-token/s training run with a 10$\times$ canonical tokens-to-target reduction is measured as 10M raw-corpus-equivalent tokens/s against the same target capability.

\begin{table*}[t]
\caption{Core Definitions}
\label{tab:definitions}
\centering
\footnotesize
\begin{tabularx}{\textwidth}{p{0.22\textwidth}X}
\toprule
Term & Definition \\
\midrule
Canonical profile & A governed language-role, layout, naming, dependency, generated-boundary, data-truth, and validation configuration for a software class. The primary app profile is a \canonprofile{}: one maintained backend/core lane, one typed product-surface lane, one browser/product-shell lane, one durable-data lane, one schema/contract lane, and one build/deploy/proof envelope. Systems software uses approved secondary profiles or governed primitives. \\
Canonical port & A behavior-preserving or behavior-dispositioned transformation from a human artifact into a canonical profile. \\
Canonical training object & A machine-readable bundle: provenance, license status, original source, port disposition, behavior evidence, canonical artifact, rejected invalid paths, and validation results. \\
Human-code entropy & Conditional representational uncertainty after behavior is fixed: the languages, frameworks, layouts, names, CI rituals, dependency choices, repository policies, and proof routes that vary without changing supported behavior. \\
Agent sprawl & The inference-time expansion of context reads, reasoning tokens, tool calls, architecture discovery, invalid edits, proof-loop failures, and review burden caused by unconstrained human repositories. \\
Canonical behavior cell & A named, versioned, proof-carrying software primitive: stable interface, schema, policy model, generated code, fixtures, tests, migration obligations, security negatives, repair memories, and known failure modes. \\
Semantic patch cell & A typed, governed change archetype such as add field, migrate nullable to required, add permission edge, rotate secret, add idempotency key, split table, add audit trail, or upgrade dependency safely. \\
Proof lane & A standard local or CI validation route: exact command, environment, required artifacts, log schema, and acceptance rule. \\
Proof-carrying change object & The structured unit of accepted change: intent, affected cells, typed diff, schema diff, migration diff, test delta, security delta, proof obligations, lane receipts, rollback plan, and provenance/license metadata. \\
Reasoning digest & A compact, versioned summary of known architecture, generated zones, proof routes, repair strategies, failure modes, rejected plans, compatibility ghosts, and prior proof objects for a repository, profile, or cell. \\
Constrained edit grammar & The machine-checkable set of legal files, typed edits, schema operations, migration forms, dependency updates, proof lanes, and generated-zone protections available to an agent. \\
Behavior intermediate representation & A product-level IR for entities, state machines, policies, permissions, effects, contracts, migrations, observability, proof obligations, and runtime constraints; source code is one generated projection. \\
Software genome & The mined atlas of recurring behavior families, variants, invariants, tests, proofs, provenance, repair memories, and generated projections across raw software history. \\
Minimum viable novelty & The irreducible product intent, domain fact, policy tradeoff, external constraint, or novel algorithm that remains after routine behavior, architecture, tests, proofs, dependencies, and deployment are generated or inherited. \\
Canonical repository policy & A governed configuration for CI, branch protection, review, secrets, releases, ownership, generated zones, dependency updates, and deployment permissions. \\
Behavior-equivalence class & The set of implementations that expose the same observable behavior, interfaces, security invariants, and migration guarantees. \\
\mfdl{} & Minimum Functional Description Length: the shortest canonical specification plus proofs plus renderer needed to produce a working system. \\
Cost per verified correct change & Total dollars, source/context/reasoning/action tokens, tool calls, wall time, failed loops, validation runs, and human review needed to produce an accepted proof-carrying behavior-preserving change. \\
Dense-active infrastructure cost/token & Accelerator-side serving cost per generated or processed token under a specified active-parameter regime, context length, batch size, memory system, precision, and quality target; dense and MoE baselines must be reported separately. \\
\bottomrule
\end{tabularx}
\end{table*}

The 40$\times$--150$\times$ range belongs to effective token-space reduction ($R_{\text{entropy}}$); the 1,000$\times$--100,000$\times$ range belongs to action-space pruning ($R_{\text{action}}$); the 30$\times$--150$\times$ central range belongs to training-token efficiency ($R_{\text{train}}$); reasoning-token, retry-loop, invalid-edit, verified-change cost, and dense-active cost/token reductions belong to separate denominators ($R_{\text{reason}}$, $R_{\text{retry}}$, $R_{\text{invalid}}$, $R_{\text{cost}}$, $R_{\text{cost/token}}$). These do not multiply cleanly. Many overlap. The composite effect must be measured on paired raw/canonical corpora and reported as cost per verified correct change.

\begin{table*}[t]
\caption{Claim Status and Required Evidence}
\label{tab:claim-status}
\centering
\footnotesize
\begin{tabularx}{\textwidth}{p{0.22\textwidth}p{0.16\textwidth}X X}
\toprule
Claim & Status & Current evidence & Required evidence \\
\midrule
Canonical trajectories are learnable & \measured & QLoRA convergence on 64,088 translated trajectories and zero measured forbidden-language markers. & Replication across model sizes, held-out repositories, stronger profile-validity checks, and task success. \\
Canonical repositories reduce same-model search cost & \targetclaim & Substrate design: file grammar, proof lanes, generated zones, and reasoning digests. & Paired raw/canonical tasks with identical model, issue lineage, hidden tests, and cost ledger. \\
Behavior preservation is adequate for training & \centralclaim & Dossier and preservation-tier design. & Large paired ports with tiered tests, traces, contracts, fuzzing, differential replay, human review, and security/migration evidence. \\
Training-token reduction & \centralclaim & Compression theory plus adjacent data-quality evidence. & Paired scaling curves to target accepted-change performance. \\
Behavior-genome economics & \moonshotclaim & No direct measurement yet; proposed substrate architecture. & Cell coverage census, mature foundry, runtime negative memory, and verified-change cost curves after amortization. \\
\bottomrule
\end{tabularx}
\end{table*}

\begin{table}[t]
\caption{What This Paper Does Not Claim}
\label{tab:notclaimed}
\centering
\footnotesize
\begin{tabularx}{\columnwidth}{p{0.28\columnwidth}X}
\toprule
Not claimed & Boundary \\
\midrule
Universal stack destiny & The primary stack is a versioned product-app profile, not a permanent answer for all domains. \\
Automatic equivalence & Tests, traces, and contracts are evidence; only proof-backed cells get proof-backed claims. \\
Literal 100$\times$ source shrink & Large ranges refer to specific denominators such as action space or routine-domain tokens-to-target. \\
Free canonicalization & Foundry, provenance, review, and verification costs must be amortized and included. \\
Human code lacks value & Human artifacts carry behavior, edge cases, incidents, and product judgment; only accidental representation is targeted. \\
\bottomrule
\end{tabularx}
\end{table}


%% file: sections/02a_correct_change_information_theory.tex
\section{Correct-Change Information Theory}
\label{sec:correct-change-info}

The missing theoretical move is to stop treating code as the primary object. The primary object is a verified behavior change. A repository is only one historical encoding of the behavior, tests, policies, provenance, and operational memory needed to accept that change.

Let $x$ be a raw repository state, $i$ an issue or requested change, $a_{1:k}$ an edit/tool trajectory, and $\Delta$ the accepted change object. A change is correct only under an oracle that includes behavior, tests, security, migration safety, provenance, and reviewer acceptance:
\begin{equation}
  \oracle(x,i,a_{1:k}) \rightarrow \{\textsc{accept},\textsc{reject},\textsc{escalate}\}.
\end{equation}
The canonical map $\canonmap$ is useful only when it preserves the information relevant to $\Delta$ while deleting information irrelevant to acceptance:
\begin{align}
  I(\canonmap(x);\Delta \mid i,\oracle)
    &\approx I(x;\Delta \mid i,\oracle), \\
  H(\canonmap(x)\mid \Delta,i,\oracle)
    &\ll H(x\mid \Delta,i,\oracle).
\end{align}
This is the paper's central learning-theoretic claim. The model should see less accidental entropy without losing the facts needed to make the right change.

\subsection{Accidental Representation Tax}

For a behavior target $y$, define the behavior-equivalence class
\begin{equation}
  \mathcal{E}_{y,\tau}=\{x : \oracle(x,y)\ge \tau\}.
\end{equation}
Raw software often contains many members of this class: different languages, frameworks, layouts, naming schemes, migration styles, tests, CI dialects, dependency wrappers, and repository policies. Canonicalization selects a smaller governed set $\mathcal{C}_{y,\tau}$ plus audited exceptions. The accidental representation tax is:
\begin{equation}
  \mathrm{ART}(y,\tau)=\log |\mathcal{E}_{y,\tau}|-\log |\mathcal{C}_{y,\tau}|.
\end{equation}
The exact cardinalities are not directly observable for realistic software, but the tax can be estimated by source-token counts, identifier entropy, AST-pattern entropy, file-path entropy, dependency entropy, proof-lane entropy, repository-policy entropy, and model perplexity under paired raw/canonical corpora.

This reframes the largest claim. The model is not merely learning fewer bytes. It is learning fewer behavior-equivalent encodings. That is why clone studies, naturalness results, software product-line engineering, compiler IRs, e-graphs, proof-carrying code, and library-learning systems all matter: they show different historical paths toward the same idea that repeated behavior should be represented by reusable structure rather than re-authored text~\cite{dejavu2017,hindleNaturalness,clementsProductLines,lattnerLLVM,lattnerMLIR,eggEgraphs,proofCarryingCode,dreamcoder}.

\subsection{Correct-Change Search Work}

A coding agent pays for branching. At each step $t$, it faces legal or perceived choices $A_t$: files to inspect, commands to run, edit locations, migration strategies, dependency changes, proof lanes, and repair paths. Raw repositories inflate $|A_t|$ with local folklore. Canonical repositories should shrink the legal set and reject invalid moves before they become failed trajectories. A crude lower bound for search work is:
\begin{equation}
  W_{\mathrm{search}} \propto \sum_{t=1}^{k}\log |A_t| + \sum_{j=1}^{m} \rho_j F_j,
\end{equation}
where $F_j$ are failed verification or review loops and $\rho_j$ are their token/tool/human penalties. Canonicalization wins when it reduces both the action entropy and the expected number of expensive failures.

This is why the strongest near-term experiment is same-model raw/canonical ablation. It does not require a new model or a full training run. It asks whether the substrate itself reduces search work: fewer files opened, fewer tool calls, fewer invalid edits, fewer failed proof lanes, fewer review comments, and lower dollars per accepted change.

\subsection{Minimum Functional Description Length}

Minimum Functional Description Length is the product-software analogue of MDL: not the shortest text file, but the shortest behavior description plus evidence plus renderer that satisfies the declared oracle. The practical objective is:
\begin{align}
\operatorname{MFDL}_{\mathcal{S},\tau}(y)
  ={}& \min_{z\in\mathcal{Z}_{\mathcal{S}}}
  \Bigl(|z|+\lambda|\Pi(z)|+\mu|\mathcal{R}_{\mathcal{S}}|
  +\nu|G(z)|\Bigr) \notag\\
& \text{s.t.}\quad
\oracle(\mathcal{R}_{\mathcal{S}}(z),y)\ge\tau.
\end{align}
Here $z$ is the canonical behavior object, $\Pi(z)$ its proof/evidence bundle, $\mathcal{R}_{\mathcal{S}}$ the renderer into source, tests, migrations, CI, documentation, observability, and deployment artifacts, and $G(z)$ the governance/provenance burden. This extra governance term matters: compression that drops attribution, opt-out state, security evidence, or compatibility ghosts is not a substrate win.

\begin{table*}[t]
\caption{From Code Imitation to Correct-Change Information}
\label{tab:information-theory-spine}
\centering
\footnotesize
\begin{tabularx}{\textwidth}{p{0.18\textwidth}X X}
\toprule
Layer & Raw-model burden & Canonical information object \\
\midrule
Text/source & Learn many strings for same behavior. & One governed rendering plus audited exceptions. \\
Repository & Infer local architecture, generated zones, and proof commands. & File grammar, generated-zone manifest, proof lanes, repository-policy object. \\
Behavior & Re-implement auth, lifecycle, billing, upload, search, jobs, webhooks, audit, observability. & Versioned behavior cells with parameters, tests, proofs, and negative cases. \\
Change & Invent raw diffs and hope tests catch mistakes. & Typed semantic patch cell with preconditions, generated deltas, proof obligations, rollback. \\
Evidence & Reviewer reconstructs intent from a diff and logs. & Proof-carrying change object with receipts, provenance, security delta, migration replay, review rubric. \\
Memory & Incidents and failed patches disappear into history. & Runtime negative memory: forbidden plans, compatibility ghosts, regression generators, patched cells. \\
\bottomrule
\end{tabularx}
\end{table*}

\textbf{Falsification:} this information theory fails if canonical artifacts preserve no more mutual information about accepted changes than raw repositories, or if they reduce entropy only by deleting behavior needed for hidden tests, product acceptance, security, migration safety, or provenance.


%% file: sections/03_human_code_entropy.tex
\section{The Human-Code Entropy Problem}

Human code is chaotic because human software production is chaotic. Teams choose languages for hiring markets, deployment constraints, fashion, deadlines, legacy compatibility, framework momentum, and personal preference. They choose continuous-integration systems and workflow YAML by copy-paste. They choose names under time pressure. They create folders that reflect organization charts and local arguments. They copy data transfer objects instead of generating contracts. They hand-edit generated artifacts. They split logic across services because the team split. They preserve test suites that prove mocks, not behavior. They keep dependencies because updating them breaks a release.

This is not an argument that human software history lacks value. The public and private code record contains product judgment, edge cases, incident responses, migration scars, security repairs, tests, failure histories, and operational lessons that a canonical substrate must preserve or disposition. The target is narrower and more aggressive: remove optional representation that remains after behavior is fixed, while carrying forward the evidence that explains why the behavior matters.

None of this is rare. GitGuardian found 28.65 million new hardcoded secrets added to public GitHub commits in 2025, a 34\% year-over-year increase~\cite{gitguardian2026}. Synopsys reported that 84\% of assessed codebases contained open-source vulnerabilities, 74\% contained high-risk vulnerabilities, and 91\% used components ten or more versions behind the current version~\cite{synopsysOSSRA2024}. GitHub Actions studies find exactly the same problem in automation: workflow complexity, heterogeneity, and compliance drift are now research subjects~\cite{ghaHeterogeneity2025}; workflow files change continuously across large repository samples~\cite{ghaEvolution2026}; security-practice adoption remains uneven even in a dominant CI/CD platform~\cite{ghaSecurityNDSS2026}. Travis CI evidence is similar: 3.7 million jobs across 1,276 projects showed noisy and heterogeneous build outcomes, including misleading pass/fail signals~\cite{travisNoise2018}, and 9,312 Travis-using projects exhibited detectable CI feature misuse~\cite{travisMisuse2020}. OpenSSF Scorecard exists because open-source projects do not consistently use secure practices~\cite{openssfScorecard,scorecardPaper}. METR's 2025 randomized trial found experienced open-source developers were 19\% slower with early-2025 AI tools on mature familiar projects, while METR's 2026 update emphasizes that adoption and task-selection effects make real productivity measurement difficult~\cite{metr2025,metr2026}.

These findings do not prove every human program is broken. They prove the stronger substrate point: human software lacks a single enforced shape. The code may run, but the representation is filled with optionality that does not encode product behavior. For an agent, optionality is not freedom; it is branching factor. If there are 20 plausible files, five plausible CI fixes, three dependency-update conventions, four migration styles, and several local naming dialects, the agent must spend tokens and tool calls discovering local folklore before changing behavior. This is \textbf{agent sprawl}: the model becomes a repository cartographer, build engineer, migration reviewer, dependency archaeologist, security analyst, and code reviewer before it can safely edit behavior. A canonical substrate removes that burden by deleting opinions from the supported path.

\begin{table*}[t]
\caption{Human-Code Variation That Bloats the Training Target}
\label{tab:chaos}
\centering
\footnotesize
\begin{tabularx}{\textwidth}{p{0.18\textwidth}X X}
\toprule
Variation class & Human-code reality & Canonical target \\
\midrule
Language sprawl & Same backend behavior appears across many service, scripting, and systems languages. & One primary language per role, with secondary profiles for true infrastructure primitives. \\
Framework drift & Express, FastAPI, Spring, Django, Rails, Laravel, Next, Remix, custom RPC, custom queues. & One canonical service grammar and one canonical UI/product grammar per profile. \\
Naming drift & User/account/customer/member; manager/service/handler/repo/dao; vague helpers and utils. & Controlled vocabulary tied to role, boundary, and domain. \\
Layout drift & Arbitrary folders, mixed generated and source files, stale migrations, implicit ownership. & Enforced file grammar, generated boundaries, and ownership maps. \\
CI/build drift & GitHub Actions, GitLab CI, Travis, Circle, custom runners, local env assumptions, undocumented gates, skipped or flaky tests. & One proof-lane grammar with reproducible commands, artifacts, permissions, and failure classes. \\
Dependency drift & Unpinned libraries, abandoned packages, vulnerable transitive dependencies, risky workflows. & Governed dependency closure, upgrade lanes, and scored exceptions. \\
Data truth drift & Data-transfer-object forks, direct queries, app-owned durable truth, duplicated schema definitions. & Governed durable-data truth, generated contracts, typed adapters, migration policy. \\
Repository-policy drift & Branch protection, code owners, required checks, release permissions, environments, secrets, and deployment credentials vary per team. & One repository-policy object with generated GitHub/GitLab settings and auditable exceptions. \\
Agent ambiguity & Model guesses where to edit and what proves completion. & Agent-readable scope, canonical commands, valid repair paths, and no-edit generated zones. \\
\bottomrule
\end{tabularx}
\end{table*}

This variation is not free. Every equivalent way to express a behavior becomes probability mass the model must learn. Every framework convention becomes a routing burden. Every naming style becomes lexical entropy. Every CI dialect becomes an operational trap. Every ungoverned repository setting becomes hidden state. The frontier model pays for all of it.

Agent sprawl is not solved by a larger context window alone. A larger window lets the model carry more accidental state; it does not remove the accidental state. The canonical countermeasure is to move repeated reasoning into substrate law: file grammar tells the model where behavior lives, generated-zone manifests tell it where not to edit, proof lanes tell it what proves completion, behavior cells identify the recurring primitive, semantic patch cells define legal change operations, reasoning digests summarize known architecture and failure modes, and proof-carrying change objects make review evidence explicit.


%% file: sections/04_assumption_a_profile.tex
\section{Assumption A: Versioned Product Profiles and Canonical Programming}

\textbf{Assumption A: For each supported software class, there should be one governed way to express, test, secure, deploy, and repair each concern.} The primary agent-first canonical profile targets web, product, SaaS, backend, and data applications---not all software. We denote the current profile $P_{\mathrm{app}}^{2026}$ as a \canonprofile{}: one maintained backend/core lane, one typed product-surface lane, one browser/product-shell lane, one durable-data lane, one schema/contract lane, and one build/deploy/proof envelope, plus repository policy and observability. Systems code, kernels, compilers, databases, browsers, GPU kernels, embedded, native mobile, and high-performance computing use \emph{secondary canonical profiles} or \emph{governed primitives}. The thesis is not implementation monoculture. The thesis is that human preference should not decide the shape of routine product software.

The profile is replaceable. The invariant is governed role ownership, generated truth, proof lanes, constrained edits, provenance, and observability:
\begin{equation}
\begin{aligned}
P_{\mathrm{app}}^{t+1} =
\operatorname{Govern}(&P_{\mathrm{app}}^{t}, E_{\mathrm{security}}, E_{\mathrm{agent}},\\
&E_{\mathrm{performance}}, E_{\mathrm{ecosystem}}, E_{\mathrm{migration}}).
\end{aligned}
\end{equation}
Profile components change only when evidence shows lower accepted-change cost, stronger verification reliability, better security, or better ecosystem support after transition cost.

Canonical does not mean a single unchecked implementation. It means one governed interface, multiple independently validated implementations where correlated failure would be catastrophic. Critical cells require independent implementations where feasible, adversarial review, conformance suites, canary rollout, version pinning, rollback, deprecation protocols, provenance records, and profile-specific exceptions. Unlimited drift is rejected; audited diversity is retained.

The canonical standard has six hard properties:

\textbf{First, profile roles are fixed.} The profile assigns one supported lane to backend/core/tooling, one lane to typed product surfaces, one lane to the browser product shell, one lane to durable data, one lane to deployment envelopes, and one lane to generated contracts. The concrete technologies are versioned profile choices, not the thesis. The invariant is that routine changes occur against explicit role ownership rather than local framework folklore.

\textbf{Second, file grammar is enforced.} Source paths, generated paths, tests, migrations, contracts, adapters, and product surfaces live in predictable places.

\textbf{Third, naming is canonical.} Variable and module names encode stable roles: actor, resource, command, event, policy, adapter, projection, migration, contract, and view.

\textbf{Fourth, generated boundaries are sacred.} Contracts generate clients, schemas, test fixtures, and adapters. Generated outputs are not hand-edited.

\textbf{Fifth, continuous integration, build, and proof commands are standard.} Every change maps to known local commands, CI jobs, permissions, logs, artifacts, and failure classes.

\textbf{Sixth, repository policy, dependency policy, and data truth are governed.} One durable-data lane owns persistent truth; dependencies are pinned, owned, scored, and upgraded through canonical lanes; branch protection, required checks, code owners, release gates, environment secrets, and deploy permissions are generated from one repository-policy object.

\begin{table*}[t]
\caption{Canonical Standard as a Compression Language}
\label{tab:standardcompression}
\centering
\footnotesize
\begin{tabularx}{\textwidth}{p{0.18\textwidth}X X}
\toprule
Standard element & Human-code space removed & Training effect \\
\midrule
Limited language roles & Equivalent business behavior expressed across many backend, frontend, script, and migration languages. & Fewer grammars and lower cross-language transfer burden. \\
Canonical file grammar & Local folder myths, mixed generated/source files, arbitrary helpers, unclear ownership. & Repository layout becomes predictable context instead of hidden state. \\
Controlled vocabulary & Synonym drift, vague service names, local abbreviations, duplicated domain labels. & Lower lexical entropy and easier long-context retrieval. \\
Generated boundaries & Data-transfer-object forks, hand-edited clients, duplicated schema truth, stale fixtures. & One source of truth replaces many invalid patch paths. \\
Strict build/test commands & README folklore, local scripts, skipped tests, inconsistent CI gates. & Model learns standard completion actions instead of guessing rituals. \\
Governed dependencies & Unpinned packages, stale transitive libraries, unsafe workflow permissions. & Dependency updates become canonical maintenance tasks. \\
Data truth policy & Direct writes, app-owned durable truth, duplicate validation logic, unsafe migrations. & Database and migration repairs become standard forms. \\
Repository policy & Hand-configured GitHub/GitLab settings, inconsistent branch protection, unscoped workflow permissions, unclear owners. & Review, merge, release, and secret-handling become generated infrastructure. \\
\bottomrule
\end{tabularx}
\end{table*}

The canonical standard is therefore a compression language for software. It compresses by removing representational degrees of freedom that do not change the product. If two repositories implement the same account-management route with eight framework stacks, four CI systems, three naming conventions, and incompatible deploy policies, a broad model pays for all of them. A canonical model pays for the behavior once, plus the governed exceptions that matter.

\textbf{Falsification:} Assumption A fails if paired raw/canonical corpora do not reduce source tokens, identifier entropy, AST-pattern entropy, path grammar, perplexity, or action branching on target software classes.


%% file: sections/05_assumption_b_foundry.tex
\section{Assumption B: Governed Foundry and Port Dispositions}

\textbf{Assumption B: Every available artifact in the training set can be assigned a governed port disposition, and the cost of porting amortizes across repeated training runs, serving savings, and downstream reuse.}

Full coverage means every artifact enters the foundry. The claim is not that every repository becomes one product-app implementation. The claim is that every artifact is either transformed into a canonical training object or explicitly excluded with a reason that remains useful for governance, evaluation, or negative training. The maturity ladder is staged: first \emph{disposition coverage}, then \emph{behavior-fixture extraction}, then \emph{full canonical porting} only where value, license status, and preservation evidence justify the cost.

\begin{table*}[t]
\caption{Governed Port Dispositions}
\label{tab:dispositions}
\centering
\scriptsize
\begin{tabularx}{\textwidth}{p{0.17\textwidth}X X}
\toprule
Disposition & Meaning & Training use \\
\midrule
\texttt{transform-permitted} & License, provenance, security, and behavior evidence permit a canonical artifact or full port. & Canonical source, behavior object, transformation trace, accepted patch examples. \\
\texttt{secondary-profile} & Artifact belongs to systems, mobile, embedded, scientific, or other non-product-app profile. & Profile-specific training and cross-profile interface examples. \\
\texttt{governed-primitive} & Foundational component is retained behind stable wrappers, tests, versions, and ownership. & Dependency, wrapper, compatibility, upgrade, and conformance examples. \\
\texttt{spec-only} & Source cannot be transformed, but public behavior/specification can be represented independently. & Abstract contracts, interface facts, and conformance targets. \\
\texttt{fixture-only} & Tests, traces, issues, API examples, or migration fixtures are usable without source projection. & Behavior fixtures and weak-oracle warnings. \\
\texttt{negative-only} & Artifact teaches a failure mode, vulnerability, bad migration, or rejected pattern. & Negative examples, refusal cases, and repair memories. \\
\texttt{metadata-only} & Only aggregate metadata, disposition statistics, or provenance facts can be retained. & Governance metrics and exclusion statistics. \\
\texttt{quarantine} & Artifact appears unsafe, malicious, poisoned, undispositioned, or provenance-risky. & No positive training; security analysis and exclusion receipt only. \\
\texttt{license-excluded} & License, attribution, opt-out, or derivative-work risk blocks inclusion. & No positive training artifact; removal propagation and audit receipt. \\
\bottomrule
\end{tabularx}
\end{table*}

Every canonical training object is an auditable data product, not a loose rewrite. Its minimum schema includes source identity, license, profile, disposition, transformation trace, behavior evidence, security evidence, canonical artifact, known loss, and training-use policy.
Source identity includes repository URL, commit, content hash or Software Heritage identifier when available, opt-out/removal state, and original artifact digest. License and supply-chain evidence should use existing machine-readable standards where possible: SPDX for license/provenance and software-bill-of-material information, CycloneDX for broader bill-of-materials and cyber-risk metadata, SLSA/in-toto-style attestations for build and transformation steps, OpenTelemetry semantic conventions for runtime evidence, and policy-as-code systems such as Open Policy Agent for inclusion and release rules~\cite{spdxOverview,cyclonedx,openTelemetrySemconv,openPolicyAgent,slsa,intoto}.

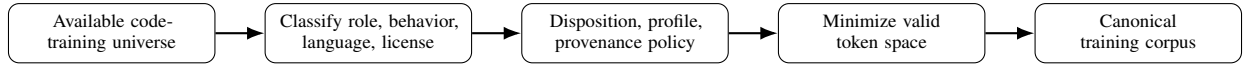
\begin{figure*}[t]
\centering
\scriptsize
\begin{tikzpicture}[
  node distance=0.65cm,
  box/.style={draw, rounded corners, align=center, minimum height=0.75cm, text width=2.45cm, inner sep=4pt},
  arr/.style={-Latex, thick}
]
\node[box] (raw) {Available code-training universe};
\node[box, right=of raw] (classify) {Classify role, behavior, language, license};
\node[box, right=of classify] (port) {Disposition, profile, provenance policy};
\node[box, right=of port] (minimize) {Minimize valid token space};
\node[box, right=of minimize] (train) {Canonical training corpus};
\draw[arr] (raw) -- (classify);
\draw[arr] (classify) -- (port);
\draw[arr] (port) -- (minimize);
\draw[arr] (minimize) -- (train);
\end{tikzpicture}
\caption{Governed full-corpus foundry. The corpus is not filtered down to clean examples; every artifact receives provenance-aware disposition, and only qualified artifacts become positive canonical training objects.}
\label{fig:porting}
\end{figure*}

\textbf{Falsification:} Assumption B fails if a representative corpus cannot be assigned governed dispositions, if license/provenance metadata cannot be preserved through transformation and removal propagation, if behavior fixtures cannot be extracted for high-value classes, or if porting cost does not amortize across at least 3$\times$ the initial foundry investment.


%% file: sections/06_assumption_c_behavior_preservation.tex
\section{Assumption C: Behavior Preservation Under Declared Contracts}

\textbf{Assumption C: Canonical ports can preserve, disposition, or explicitly reject behavior at a confidence level high enough for training.} This is the hardest assumption in agent-first canonical code. A canonical rewrite that silently deletes edge cases is not compression; it is corruption. Fully automatic behavior equivalence for arbitrary programs is not available in general; Rice's theorem is the theoretical warning label on any claim that a tool can decide nontrivial semantic properties for all programs~\cite{riceTheoremMIT}. The foundry therefore treats behavior evidence as part of the training object, not a best-effort annotation.

Every canonical training object carries a preservation and divergence dossier: original tests, canonical tests, differential traces, API compatibility checks, UI replay where applicable, migration replay, security-negative cases, performance budgets for behavior that depends on latency or resource use, provenance/license records, accepted incompatibilities, compatibility ghosts, and rejected port paths. A canonical port is never behavior-preserving merely because tests pass; it is behavior-preserving only within a declared evidence envelope.

The output is not binary pass/fail. It is graded behavior confidence for original artifact $o$ and canonical artifact $z$:
\begin{equation}
  B(o,z)=\sum_{i\in\mathcal{E}} w_i e_i(o,z),
  \qquad 0\le e_i\le 1,\quad \sum_i w_i=1.
\end{equation}
Here $\mathcal{E}$ includes test agreement, differential execution coverage, migration and data-invariant replay, security-negative coverage, UI/API replay coverage, human acceptance for high-value cases, performance/SLO agreement where applicable, and provenance/license completeness.

\begin{table*}[t]
\caption{Behavior Preservation Dossier}
\label{tab:behaviorpreservation}
\centering
\footnotesize
\begin{tabularx}{\textwidth}{p{0.20\textwidth}X X}
\toprule
Evidence lane & What it catches & Canonical training use \\
\midrule
Original tests & Known intended behavior and regression expectations. & Positive evidence and weak-oracle metadata. \\
Differential traces & Runtime behavior across original and ported artifacts. & Equivalence scoring and edge-case discovery. \\
Migration replay & Data-shape, rollback, lock-budget, and invariant failures. & SQL/migration proof examples. \\
Security negatives & Auth bypass, injection, secret leakage, unsafe defaults. & Labeled bad paths for refusal and repair training. \\
UI/API replay & Browser, CLI, and API compatibility surfaces. & Product behavior fixtures. \\
Human acceptance & High-value cases where automated tests are insufficient. & Gold labels and accepted-incompatibility manifests. \\
Provenance/license & Legal and origin constraints. & Inclusion, exclusion, and attribution metadata. \\
\bottomrule
\end{tabularx}
\end{table*}

Compatibility ghosts are first-class. If downstream users depend on behavior that looks accidental---ordering, timing, error text, weak validation, legacy API quirks, migration side effects, or undocumented defaults---the foundry must either preserve it, encode it as a compatibility fixture, or record an accepted incompatibility signed by the relevant owner. Calling it accidental is not enough to delete it.

The output is a preservation tier, not a blanket declaration of equivalence:

\begin{table*}[t]
\caption{Canonical Preservation Tiers}
\label{tab:preservationtiers}
\centering
\footnotesize
\begin{tabularx}{\textwidth}{p{0.10\textwidth}p{0.22\textwidth}X X}
\toprule
Tier & Name & Meaning & Training use \\
\midrule
P0 & Rejected / metadata-only & Unsafe, unlicensed, malicious, undispositioned, or behavior evidence is too weak. & Exclusion, quarantine metadata, negative examples only. \\
P1 & Syntactic & Builds, formats, typechecks, or parses, but no behavior claim is made. & Low-weight syntax/profile examples only. \\
P2 & Interface/test & Public interfaces are represented and original/canonical tests or fixtures pass. & Weak positive training with explicit weak-oracle label. \\
P3 & Trace/contract & API, schema, migration, policy, compatibility ghosts, and negative cases match declared contracts and replay traces. & Standard product-app training target. \\
P4 & Property/fuzz/differential & Generated, fuzzed, property-based, symbolic, mutation-tested, or production-derived inputs exercise critical paths. & Higher-confidence port and edge-case curriculum. \\
P5 & Proof-backed & A formal or mechanized proof discharges specified invariants or semantic preservation obligations. & Certified cell, compiler/profile primitive, or critical dependency. \\
\bottomrule
\end{tabularx}
\end{table*}

The weak-oracle problem is explicit. Tests are evidence, not proof. SWE-bench-style patch evaluation can be distorted by insufficient tests and plausible-but-wrong patches; UTBoost found insufficient test cases and erroneous patches previously labeled as passed in SWE-bench-family evaluation~\cite{utboost}. Property-based testing, differential testing, compiler fuzzing, mutation testing, and symbolic execution are stronger evidence lanes, but they remain scoped evidence rather than universal equivalence~\cite{quickcheck,csmith,mutationTesting,kleeOSDI}. CompCert shows that semantic preservation can be a compiler contract rather than a slogan~\cite{compcertManual}. seL4 shows that proof-backed systems components are possible for security-critical kernels, while also making clear how much engineering work such proofs require~\cite{sel4Proofs}. The canonical foundry should use these methods where their cost is justified and reject overclaiming elsewhere.

This section is intentionally early because behavior preservation is the trust anchor. The authorial claim that all human code is improvable does not mean all human code is disposable. It means every human artifact can be made more canonical if the behavior worth keeping is first identified, tested, replayed, and carried forward.

\textbf{Falsification:} Assumption C fails if behavior confidence remains low, human review rejects core ports, security/migration regressions rise, or important edge behavior is routinely lost during canonicalization.


%% file: sections/07_assumption_d_behavior_cells.tex
\section{Assumption D: Canonical Behavior Cells}

\textbf{Assumption D: The highest-reuse software behaviors can be collapsed into certified, reusable behavior cells, and these cells can eventually absorb 70\%--90\% of routine product code if a behavior-cell census validates coverage.}

This is the breakthrough beyond canonical stack alignment. If the model is still writing 80 functions for create/read/update/delete resource behavior, authentication, forms, validation, errors, pagination, tests, clients, fixtures, and migrations, we are only standardizing the mess. If the model emits a \emph{cell declaration and parameters}, we are removing entire classes of source tokens, legal edits, proof obligations, and review questions from the task.

A canonical behavior cell is not a library. It is a reusable behavior unit with a stable interface, schema, policy model, generated projections across the full stack (service, typed client, product UI, durable-data migration, tests, fixtures, observability, and docs), known failure modes, security negatives, repair memories, compatibility ghosts, and proof obligations. Change itself also becomes cell-shaped: common edits such as add field, migrate nullable to required, add permission edge, rotate secret, add idempotency key, upgrade dependency safely, split table, and add audit trail become semantic patch cells rather than bespoke diff inventions.

For routine domains, behavior cells are the main path from canonical code to \mfdl{}. The agent should not decide from scratch how authorization, audit trails, pagination, idempotency, upload signing, or migration rollback work in every repository. It should select a certified cell, supply parameters, apply a semantic patch cell when behavior changes, and return a proof-carrying change object with receipts.

The 70\%--90\% number is a target, not a measured result in this paper. It must be earned by a behavior-cell census that measures coverage by accepted changes, source tokens, AST nodes, runtime traces, issue tickets, security-sensitive paths, review burden, and production incident classes. A cell is useful only if it reduces verified-change cost without hiding required variation.

\begin{table*}[t]
\caption{Behavior Cells: From Human-Code Mess to Canonical Primitives}
\label{tab:cells}
\centering
\footnotesize
\begin{tabularx}{\textwidth}{p{0.37\textwidth}p{0.22\textwidth}X}
\toprule
Human-code mess & Canonical cell & Why it compresses \\
\midrule
Rails users controller, Express users route, FastAPI users service, Spring user repository & \texttt{resource<User>} & One verified cell replaces thousands of reimplementations. \\
Custom login/session/token/password-reset/multi-factor code & \texttt{auth.session} / \texttt{auth.oidc} & Repeated in every app with dangerous variation. \\
Hand-written role- or attribute-based permission checks scattered across layers & \texttt{policy.check} & Security-critical code becomes a proven primitive. \\
Custom list endpoints with ad hoc pagination/sorting/filtering & \texttt{list.page} & Massive API/UI/DB repetition becomes one cell. \\
Ad hoc form validation duplicated across frontend/API/database/tests & \texttt{schema.form.validation} & Four validation layers generated from one schema. \\
Recurring billing glue, Stripe webhook wrappers, proration logic & \texttt{billing.subscription} & Common SaaS surface with dangerous edge cases. \\
Background retries, idempotency, dead-letter queues & \texttt{job.retry.idempotent} & Same reliability logic reimplemented everywhere. \\
Object-storage upload variants, signed URLs, virus scanning & \texttt{blob.upload.signed\_url} & Security and storage complexity repeated constantly. \\
Custom webhook verification, replay, deduplication & \texttt{webhook.verify.dispatch} & Integration-heavy apps repeat this badly. \\
Audit trails, event history, compliance logging & \texttt{audit.append\_only} & Required for compliance-heavy systems. \\
Expand/contract migrations, backfills, rollback plans & \texttt{migration.expand} & High failure cost and high repetition. \\
Custom notification channels, delivery tracking, templates & \texttt{notification.send} & Repeated boilerplate across channels. \\
\bottomrule
\end{tabularx}
\end{table*}

The compression ladder shows how gains compound across the standardization levels:

\begin{table*}[t]
\caption{The Compression Ladder: From Raw Repository to Canonical Behavior Graph}
\label{tab:ladder_compression}
\centering
\footnotesize
\begin{tabularx}{\textwidth}{p{0.04\textwidth}p{0.18\textwidth}p{0.28\textwidth}X}
\toprule
Rung & Level & What changes & Measurement \\
\midrule
0 & Raw human repo & Original code & Baseline \\
1 & Canonical layout & Fixed paths, ownership, generated zones & File-search entropy, edit target count \\
2 & Canonical naming & Actor/resource/command/event/policy vocabulary & Identifier entropy \\
3 & Single primary profile & Backend/core, typed surface, product shell, durable data, contracts, build/deploy/proof & Language/framework entropy \\
4 & Contract-first generation & Schema generates clients, fixtures, validators, docs & Duplicate data-transfer-object/client drift \\
5 & Proof lanes & Fixed build/test/migrate/security commands & Failed-loop count \\
6 & Behavior cells & Authentication, resource lifecycle, billing, search, uploads, jobs as primitives & Source-token and AST-pattern reduction \\
7 & Behavior graph & Model emits behavior graph operations plus proof obligations; source code is compiled & Task-token and reasoning-token reduction \\
\bottomrule
\end{tabularx}
\end{table*}

The endgame is rung 7: models manipulate canonical software behavior graphs, and source code becomes a generated artifact. The training target moves from source text toward graph operations, semantic patches, proof obligations, and receipts. This is analogous to how LLVM created a reusable IR for program analysis and transformation~\cite{lattnerLLVM}, and how MLIR explicitly aims to reduce software fragmentation~\cite{lattnerMLIR}. Software product-line engineering has a related history: explicitly modeling commonality and variability across product variants to build reusable core assets~\cite{clementsProductLines}. Equality-saturation systems such as egg show how rewrite spaces can be represented and searched compactly when the intermediate representation is explicit~\cite{eggEgraphs}.

Cells have a lifecycle, not just a registry entry:
\begin{table}[t]
\caption{Behavior Cell Lifecycle}
\label{tab:cell-lifecycle}
\centering
\footnotesize
\begin{tabularx}{\columnwidth}{p{0.28\columnwidth}X}
\toprule
Stage & Required evidence \\
\midrule
Propose & Behavior family, interface, parameters, variants, provenance, and expected coverage. \\
Certify & Preservation tier, conformance suite, security negatives, fuzz/property lanes, rollout risk. \\
Deploy & Version pinning, generated projections, proof receipts, canary plan, rollback path. \\
Monitor & Runtime traces, incidents, performance budgets, compatibility ghosts, exploit reports. \\
Deprecate & Migration path, old-version risk, user impact, and negative-memory update. \\
Incident response & Revocation, patched cell release, downstream blast-radius scan, and postmortem fixtures. \\
\bottomrule
\end{tabularx}
\end{table}

\textbf{Falsification:} Assumption D fails if the top 500--2,000 cells cover less than 50\% of routine product-app behavior, or if cell composition introduces more complexity than it removes.


%% file: sections/08_empirical_validation.tex
\section{Preliminary Learnability and Profile-Adherence Evidence}

\textbf{Claim scoped to this section: canonical data is learnable and can enforce target-profile adherence.} There is prior evidence that code model performance benefits heavily from data quality: Arctic-SnowCoder-1.3B used 555B tokens in staged data refinement and beat or matched models trained on much larger token budgets~\cite{arcticSnowCoder}. The phi-1 work showed that high-quality ``textbook'' data can produce remarkably capable small models~\cite{phi1}.

To test whether canonical-translated data produces healthy convergence, we ran a quantized low-rank adaptation (QLoRA) fine-tuning experiment on Qwen2.5-Coder-14B-Instruct~\cite{qlora,qwenCoder} using 64,088 canonically translated agentic coding trajectories (57,688 train, 6,400 validation). The dataset consists of verified multi-turn coding sessions where an agent reads a real GitHub issue, explores the codebase, applies a fix, and passes automated tests, then translates the trajectory through the canonical porting pipeline to align with the target limited-stack profile.

Training ran for 7,106 steps on a single NVIDIA RTX~3090 (24\,GB) over approximately three days. Key hyperparameters: LoRA rank $r{=}16$, $\alpha{=}32$, learning rate $2{\times}10^{-5}$ with constant-with-warmup schedule, effective batch size $1{\times}8$ gradient accumulation, maximum sequence length 2,048 tokens, 4-bit NF4 quantization with double quantization.

\begin{table}[t]
\caption{Pilot Configuration}
\label{tab:pilotconfig}
\centering
\footnotesize
\begin{tabularx}{\columnwidth}{p{0.34\columnwidth}X}
\toprule
Field & Value \\
\midrule
Base model & Qwen2.5-Coder-14B-Instruct \\
Adaptation & QLoRA, rank 16, alpha 32, 4-bit NF4, double quantization \\
Data & 64,088 translated trajectories; 57,688 train / 6,400 validation \\
Steps & 7,106 training steps; 711 validation evaluations \\
Sequence length & 2,048 tokens \\
Hardware & Single NVIDIA RTX 3090, 24\,GB \\
Forbidden marker metric & Average presence of off-profile markers such as Python \texttt{def}, Go \texttt{func}, and Java \texttt{public class} under target-profile service prompts \\
\bottomrule
\end{tabularx}
\end{table}

\begin{figure*}[t]
  \centering
  \includegraphics[width=\textwidth]{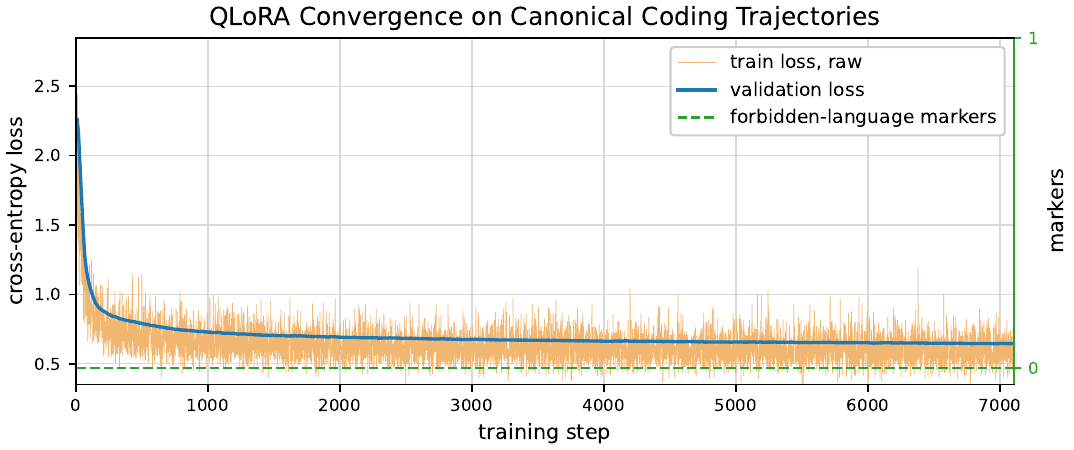}
  \caption{Raw training and validation loss convergence for QLoRA fine-tuning of Qwen2.5-Coder-14B on 64,088 canonically translated trajectories. The figure shows the real per-step training log after restart de-duplication: 7,106 training steps and 711 validation evaluations. Training loss starts at 2.707 and ends at a raw final value of 0.419. Validation loss drops from 2.261 at step 10 to 0.646 at step 7,106, with a minimum of 0.644 at step 6,780. Zero forbidden-language markers were recorded across all evaluation checkpoints. The vector figure is generated from raw per-step training logs and evaluation telemetry only.}
  \label{fig:convergence}
\end{figure*}

Figure~\ref{fig:convergence} shows the convergence curve generated from raw training telemetry. After restart de-duplication, the run contains 7,106 training steps and 711 validation evaluations. Raw training loss drops from 2.707 at step 1 to 0.419 at step 7,106. Validation loss drops from 2.261 at step 10 to 0.646 at step 7,106, closely tracking the raw training curve throughout. Two observations are significant:

\begin{enumerate}
\item \textbf{No visible train/validation divergence.} Raw training and validation losses fall together through the run, with validation ending at 0.646 after consuming 100\% of the training data. By contrast, raw open-source coding data typically requires aggressive deduplication: The Stack v2 reduced from 67.5\,TB to 32.1\,TB (52\% deduplication) and then to 2.4\,TB after quality filtering---a 96.4\% reduction.

\item \textbf{Zero forbidden-language violations.} Throughout training, evaluation checkpoints measured off-profile markers (Python \texttt{def}, Go \texttt{func}, Java \texttt{public class}, etc.) in model outputs prompted with target-profile service tasks. All 711 evaluation checkpoints recorded zero markers, indicating that canonical fine-tuning successfully steered the model toward the target profile.
\end{enumerate}

\begin{table}[t]
\caption{What the Pilot Proves and Does Not Prove}
\label{tab:pilotlimits}
\centering
\footnotesize
\begin{tabularx}{\columnwidth}{p{0.32\columnwidth}X}
\toprule
Status & Claim \\
\midrule
Shows & Canonical translated trajectories are learnable under this parameter-efficient adaptation setup. \\
Shows & The measured forbidden-language markers can be suppressed in the evaluation prompts. \\
Does not show & Behavior preservation, full stack correctness, held-out accepted-change success, or lower cost per verified change. \\
Does not show & Tokens-to-target reduction versus raw trajectories; that requires paired scaling curves from the same task lineage. \\
\bottomrule
\end{tabularx}
\end{table}

These results provide early evidence that canonical data is learnable under this parameter-efficient adaptation setup. They do not yet prove behavior preservation, cost-per-correct-change reduction, or 100B replacing 1T. Those require the full evaluation protocol described in Section~\ref{sec:eval}.


%% file: sections/09_assumption_e_compression_theory.tex
\section{Assumption E: Compression Theory and Minimum Functional Description Length}

\textbf{Assumption E: Removing accidental human degrees of freedom reduces effective representation entropy, legal action-space, and training-token demand.} Agent-first canonical code is not a source minifier. It is a canonical coding theory for software behavior. The raw corpus contains many strings for the same behavior because humans choose different languages, frameworks, folders, continuous-integration workflows, repository settings, dependency wrappers, migration rituals, and naming schemes. The canonical standard maps that equivalence class to a smaller canonical space, then constrains future edits through schemas, proof lanes, and semantic patch cells.

Let $X$ be raw human-code representations and $Z$ be canonical behavior representations. Let $Y$ be observable software behavior. The canonical standard is useful only if the canonical map preserves behavior while reducing irrelevant representation entropy:
\begin{equation}
  H(Z \mid Y) \ll H(X \mid Y).
\end{equation}
This is the formal version of ``one governed way per supported software class.'' For a supported concern, if two implementations expose the same behavior and invariants, they should not remain two equally privileged training targets. One interface becomes canonical; independent implementations, port traces, negative paths, compatibility fixtures, or excluded artifacts remain only when they carry evidence, diversity, governance, or security value.

\subsection{Compression Accounting Protocol}

All ranges in this paper are priors or targets until promoted by measurement. A range becomes measured only when the numerator, denominator, task distribution, model, evidence tier, and confidence interval are reported. For task distribution $\tau$ and target quality $q$:
\begin{align}
R_{\mathrm{source}}(\tau)
  &= \frac{T_{\mathrm{raw\ source}}(\tau)}
          {T_{\mathrm{canonical\ source}}(\tau)}, \\
R_{\mathrm{train}}(q,\tau)
  &= \frac{D_{\mathrm{raw}}(q,\tau)}
          {D_{\mathrm{canonical}}(q,\tau)}, \\
R_{\mathrm{reason}}(\tau)
  &= \frac{T_{\mathrm{reason+tool,raw}}(\tau)}
          {T_{\mathrm{reason+tool,canonical}}(\tau)}, \\
R_{\mathrm{change}}(\tau)
  &= \frac{C_{\mathrm{verified\ change,raw}}(\tau)}
          {C_{\mathrm{verified\ change,canonical}}(\tau)}.
\end{align}
Composite ratios cannot be multiplied unless a fitted causal model validates separability on held-out paired tasks. This is especially important for language collapse, framework collapse, generated truth, behavior cells, and proof lanes, which often remove the same tokens, edits, or reasoning loops.

The strongest effect is not literal token shrinkage. The strongest effect is action-space and reasoning-space collapse. A raw repository asks an agent to infer where code lives, what is generated, which tests matter, whether a migration is safe, how the CI system encodes permissions, what dependency path is acceptable, and what local naming convention should be preserved. A canonical repository turns those choices into explicit law. That law may add some metadata tokens, but it removes entire families of invalid edits, failed repair loops, and rediscovery reasoning.

\begin{table*}[t]
\caption{Canonical Reduction Denominators}
\label{tab:denominators}
\centering
\footnotesize
\begin{tabularx}{\textwidth}{p{0.22\textwidth}p{0.18\textwidth}X}
\toprule
Axis & Moonshot range & Meaning \\
\midrule
Literal source-token & \conrange{$2\times$--$8\times$ conservative} & Physical source shrinkage from generation, cells, and fewer duplicate surfaces. \\
Effective representation-space & $40\times$--$150\times$ central & Fewer valid encodings for the same supported behavior. \\
Agent action-space & \aggrange{$1{,}000\times$--$100{,}000\times$ high-end} & Fewer legal files, edits, workflows, proof lanes, migration paths, dependency choices, and generated-zone mistakes. \\
Reasoning-token space & $3\times$--$50\times$ hypothesis & Fewer planning, tool, architecture-discovery, retry, and repair tokens per accepted change. \\
Retry/review loop & $3\times$--$50\times$ hypothesis & Fewer failed tests, invalid edits, ambiguous reviews, rollback surprises, and proof reruns. \\
Training-token budget & $30\times$--$150\times$ hypothesis & Fewer examples needed to reach target canonical accuracy, conditional on paired scaling curves. \\
Cost per verified correct change & $3\times$--$100\times$ hypothesis & End-to-end cost after foundry, verification, serving, reasoning, failed-loop, and maintenance amortization. \\
\bottomrule
\end{tabularx}
\end{table*}

\begin{figure*}[t]
\centering
\scriptsize
\begin{tikzpicture}[
  x=1cm,
  y=1cm,
  every node/.style={font=\scriptsize},
  layer/.style={draw=black!42, rounded corners=1pt},
  summary/.style={draw, rounded corners=1.5pt, align=left, inner sep=4pt}
]
\def\stacklayer#1#2#3#4#5#6#7{%
  \draw[layer,fill=#7] (0,#1) rectangle (10.65,#1+0.55);
  \node[anchor=west,align=left,text width=10.2cm] at (0.18,#1+0.275) {#2};
  \node[anchor=west,align=left,text width=4.7cm] at (10.95,#1+0.275) {\conrange{#3} / #4 / \aggrange{#5}\\[-1pt]{\footnotesize #6}};
}
\node[anchor=west,font=\scriptsize\bfseries] at (0,7.05) {Cumulative substrate stack, sorted by local gain};
\node[anchor=west,font=\scriptsize\bfseries] at (10.95,7.05) {local band: \conrange{conservative} / central / \aggrange{aggressive}};
\stacklayer{6.35}{Raw human software universe}{1.0x}{1.0x}{1.0x}{baseline}{black!4}
\stacklayer{5.73}{+ Corpus hygiene: dedup, provenance, generated/vendor/secret/malware filtering}{2x}{5x}{15x}{source/corpus reduction}{black!7}
\stacklayer{5.11}{+ Reasoning digests and proof caches: known plans, failure modes, prior proof objects}{3x}{10x}{50x}{reasoning-token reduction}{black!10}
\stacklayer{4.49}{+ Proof lanes and reusable receipts: verification, rollback, migration replay, policy checks}{3x}{15x}{50x}{retry/review-loop reduction}{black!13}
\stacklayer{3.87}{+ Canonical profiles: language roles, layout, naming, repository policy, dependency law, proof lanes}{4x}{20x}{60x}{effective representation}{black!16}
\stacklayer{3.25}{+ Contract-first generated truth: schemas, clients, docs, fixtures, validators, migrations}{8x}{40x}{120x}{effective representation}{black!19}
\stacklayer{2.63}{+ Behavior cells: auth, lifecycle, search, forms, billing, uploads, jobs, webhooks, audit, observability}{15x}{75x}{250x}{routine-product representation}{black!22}
\stacklayer{2.01}{+ Constrained edit grammar and semantic patch cells}{100x}{10,000x}{100,000x}{legal-action reduction}{black!25}
\draw[layer,fill=white,thick] (0,1.28) rectangle (15.65,1.82);
\node[anchor=west,align=left,text width=15.25cm] at (0.18,1.55) {\textbf{Minimum Functional Description:} irreducible product intent, external contracts, domain invariants, novelty, and evidence.};

\node[summary,fill=black!3,draw=black!60,text width=15.15cm,anchor=north west] at (0,1.00) {
\textbf{All-layer combined envelope (effective correct-change search work, not source size or dollar cost).}\\[-1pt]
\textbf{Gross local-band product if fully independent:}
\conrange{$8.64\times10^{5}\mathrm{x}$} / $4.5\times10^{11}\mathrm{x}$ / \aggrange{$6.75\times10^{15}\mathrm{x}$}
(\textit{orientation only; not claimed}).\\[-1pt]
\textbf{Overlap-adjusted claimable envelope:}
\conrange{$10^{2}$--$10^{3}\mathrm{x}$ conservative} /
$10^{4}$--$10^{6}\mathrm{x}$ central /
\aggrange{$10^{7}$--$10^{9}\mathrm{x}$ aggressive}. The conservative band is capped because source hygiene,
canonical representation, cells, legal edits, proofs, and reasoning caches share causal mechanisms.
};
\end{tikzpicture}
\caption{Canonical Compression Cascade: From Human-Code Entropy to Minimum Functional Description. The stack starts at the 1.0$\times$ raw baseline and then adds layers sorted by conservative local gain, with ties ordered by central gain. The right column gives the local hypothesis band for each layer. The final block combines all row multipliers two ways: a gross independent-axis product for orientation, and a deliberately smaller overlap-adjusted envelope for effective correct-change search work. The local ranges are measured on different denominators and are not naively multiplied. Training time, inference cost, and verified-change cost remain separate outcome denominators.}
\label{fig:compressioncascade}
\end{figure*}

The final block in Figure~\ref{fig:compressioncascade} deliberately separates the raw independence bound from the claimable combined range. The conservative envelope is not the product of every row. It is anchored by constrained legal edits plus a small number of independent reductions in representation, retry, and reasoning work; stronger bands require evidence that those reductions remain separable on paired tasks.

Contract-first generation is the first large step toward that limit: OpenAPI and Protocol Buffers are existing examples of schema-owned surfaces that generate clients, validators, and transport bindings instead of hand-maintaining parallel representations~\cite{openapi31,protobufOverview}. Semantic patch systems such as Coccinelle and constrained generation systems such as XGrammar show complementary directions: patches and outputs can be represented as structured grammars rather than unconstrained text~\cite{coccinelle,xgrammar}.

The theoretical limit is \textbf{Minimum Functional Description Length}: a frontier over canonical specifications, evidence bundles, and renderers, not a magical globally shortest program. For substrate $\mathcal{S}$, preservation threshold $\tau$, behavior target $y$, proof/evidence bundle $\Pi$, renderer $\mathcal{R}_{\mathcal{S}}$, and observable-behavior oracle $\mathcal{O}$:
\begin{multline}
\operatorname{MFDL}_{\mathcal{S},\tau}(y)=
\min_{z\in\mathcal{Z}_{\mathcal{S}}}
\left(|z|+\lambda|\Pi(z)|+\mu|\mathcal{R}_{\mathcal{S}}|\right)\\
\text{s.t.}\quad
\mathcal{O}(\mathcal{R}_{\mathcal{S}}(z),y)\ge\tau.
\end{multline}
This is not code golf and not syntax preference. It is the end state where bespoke code appears only when the behavior is novel. Routine product behavior becomes behavior graph operations, cell composition, generated surfaces, policies, schemas, migrations, semantic patches, and proof receipts. In that limit, the model is no longer trained to imitate the long tail of human implementation choices. It is trained to operate a constrained software machine.

\textbf{Falsification:} Assumption E fails if paired raw/canonical corpora do not reduce measured source tokens, identifier entropy, AST-pattern entropy, path grammar, proof-loop count, perplexity, or action branching on target software classes.


%% file: sections/09a_behavior_ir_typed_change_algebra.tex
\section{Behavior IR and Typed Change Algebra}
\label{sec:behavior-ir}

A canonical substrate should not merely prescribe file layouts. It needs an application-level behavior intermediate representation. The behavior IR represents entities, state machines, authorization edges, external contracts, effects, idempotency requirements, migration semantics, observability, privacy zones, rollout constraints, performance envelopes, and proof obligations. The IR lowers into governed service code, typed product surfaces, durable-data operations, OpenAPI/Protocol Buffers schemas, tests, dashboards, deployment manifests, and documentation. OpenAPI and Protocol Buffers already show the value of contract-first generated projections~\cite{openapi31,protobufOverview}; the canonical substrate generalizes this from API messages to product behavior.

Routine edits should then be expressed as a typed change algebra rather than raw diffs:
\begin{equation}
\Delta = (op,params,pre,post,proofs,rollback,provenance).
\end{equation}
The operation determines the required proof lanes. For example, \texttt{AddField(User.timezone, default=UTC)} expands into schema diff, migration, API projection, form state, validation, fixtures, compatibility tests, rollout, and rollback. \texttt{AddPermissionEdge(Manager, ApproveInvoice)} expands into policy graph update, negative tests, audit event, UI affordance, threat-model delta, and authorization proof. \texttt{AddIdempotencyKey(WebhookDispatch)} expands into persistence, duplicate-replay tests, observability, retry semantics, and incident-runbook update.

\begin{table*}[t]
\caption{Typed Changes Should Compile to Proof Obligations, Not Merely Files}
\label{tab:typed-change-compiler}
\centering
\footnotesize
\begin{tabularx}{\textwidth}{p{0.19\textwidth}p{0.22\textwidth}X}
\toprule
Change operation & Canonical parameters & Compiled obligations \\
\midrule
\texttt{AddField} & entity, name, type, nullability, default/backfill & Schema diff, migration replay, generated clients, UI states, fixtures, compatibility, rollback. \\
\texttt{AddPermissionEdge} & actor, resource, action, conditions & Policy graph proof, negative tests, audit log, UI affordance, threat-model delta. \\
\texttt{SplitTable} & source, target, key, backfill, dual-write window & Data migration, read/write compatibility, rollback, observability, latency budget. \\
\texttt{AddWebhook} & provider, event, signature, retry, idempotency & Contract fixture, signature verifier, replay test, dead-letter path, alerting. \\
\texttt{UpgradeDependency} & package, version, risk class, migration notes & SBOM/provenance check, changelog constraints, compatibility tests, security scan, rollback. \\
\texttt{RotateSecret} & provider, scope, rollout window & Secret inventory, dual-read window, audit receipt, rollback, leak scan. \\
\bottomrule
\end{tabularx}
\end{table*}

This is where the paper becomes more defensible and more ambitious. A raw diff asks the reviewer to reconstruct the operation. A typed change object states the operation first and makes generated diffs subordinate evidence. Proof-carrying code established the idea that producers can ship code with checkable evidence of safety~\cite{proofCarryingCode}; CompCert and seL4 show that semantic preservation and machine-checked proofs can be practical in constrained high-value domains~\cite{compcertManual,sel4Proofs}; SLSA and in-toto show that provenance and supply-chain integrity can be standardized as artifacts~\cite{slsa,intoto}. Proof-carrying change applies the same philosophy to everyday product software: the unit of work is not a patch, but a patch plus typed intent, generated obligations, proof receipts, rollback, and provenance.

The research bet is that most product changes are not semantically arbitrary. They are drawn from a finite and learnable algebra: add field, add resource, add role, add permission, add workflow state, add integration, add notification, add idempotency, split entity, migrate invariant, upgrade dependency, add audit, expose report, tighten validation, change rollout, repair flaky proof lane. If true, agentic software engineering should focus less on unconstrained patch generation and more on recognizing, composing, proving, and amortizing these typed operations.


%% file: sections/10_assumption_i_reasoning_compression.tex
\section{Assumption I: Reasoning-Token Compression and Agent-Sprawl Collapse}
\label{sec:reasoningcompression}

\textbf{Assumption I: Canonical software compresses the reasoning process required to produce a verified correct change.} Agent sprawl is a substrate failure: when a repository does not state its own laws, teams compensate with planner agents, search agents, reviewer agents, test-fixing agents, security agents, and release agents that repeatedly rediscover the same facts. Modern coding agents do not spend only source tokens. They spend planning tokens, tool tokens, hidden or explicit reasoning tokens, file-inspection tokens, failed test loops, repair attempts, and human review attention. Reasoning-model documentation and best-practice guidance already treat reasoning effort as a controllable inference resource~\cite{openaiReasoningDocs,openaiReasoningBestPractices}. Chain-of-thought, self-consistency, ReAct, and test-time planning work show that intermediate reasoning and tool use can improve outcomes, but also make inference cost a first-class denominator~\cite{chainOfThought,selfConsistency,react,planBudgetReasoning}.

The correct-change cost model is:
\begin{equation}
\begin{split}
C_{\text{change}} ={}& T_{\text{context}} + T_{\text{reason}} + T_{\text{action}} + T_{\text{verify}} \\
&+ T_{\text{repair}} + H_{\text{review}} + A_{\text{foundry}}.
\end{split}
\end{equation}
Here $T_{\text{context}}$ is repository reading, $T_{\text{reason}}$ is planning and architecture inference, $T_{\text{action}}$ is patch generation and tool use, $T_{\text{verify}}$ is proof-lane execution and log interpretation, $T_{\text{repair}}$ is failed-loop recovery, $H_{\text{review}}$ is human review burden, and $A_{\text{foundry}}$ is amortized canonical-foundry cost.

Canonical repositories reduce these terms by compiling repeated reasoning into durable artifacts. Behavior cells encode architecture. Generated-zone manifests tell the agent where not to edit. Semantic patch cells encode common repairs. Proof lanes encode the validation sequence. Repository policy encodes review, secret, dependency, and deployment law. Reasoning digests summarize known plans, failure modes, rejected plans, compatibility ghosts, and prior proof objects. A proof-carrying change object then binds the intent, affected cells, typed diff, schema diff, migration diff, test delta, security delta, proof obligations, receipts, rollback plan, and provenance metadata.

The moonshot is therefore not that agents ``think less'' in the abstract. It is that agents stop spending tokens on repository cartography, local policy discovery, command guessing, generated-zone detection, migration-law inference, and repair-loop archaeology when the substrate can state those facts once and verify them mechanically. Structured outputs and constrained decoding are useful precedents: they move part of correctness from prompt convention into machine-checkable shape~\cite{openaiStructuredOutputs,xgrammar}. Proof-carrying code established the broader principle that executable artifacts can carry evidence checked by a consumer rather than accepted on trust~\cite{proofCarryingCode}.

The measurement unit is deliberately concrete: files opened, context tokens, planning tokens, hidden reasoning budget where observable, tool calls, invalid edits, generated-zone violations, failed proof loops, validation reruns, reviewer comments, and retries per accepted change. Closed models may hide some reasoning tokens, so field trials should report direct reasoning-token counters where available and proxy them with planning text, tool loops, wall-clock, and failed proof-lane traces where not. The conservative target is 3$\times$--10$\times$ fewer context/reasoning/tool/retry tokens per verified change. The central target is 10$\times$--100$\times$. The aggressive 100$\times$--10{,}000$\times$ band is only for routine supported work after behavior cells, semantic patch cells, proof lanes, and negative memory mature. It is not a universal inference-speed claim.

Reasoning compression is not separate from economics; it is one term in the verified-change cost equation in Assumption F. A reasoning digest only matters if it reduces measured files opened, context tokens, reasoning tokens, tool calls, invalid edits, repair attempts, reviewer comments, or wall-clock at the same accepted-change standard.

\textbf{Falsification:} Assumption I fails if, on paired raw/canonical tasks with the same broad model, canonical repositories do not reduce files opened, tool calls, reasoning tokens, invalid edits, failed verification loops, wall time, reviewer comments, or cost per accepted proof-carrying change.


%% file: sections/10_assumption_f_economics.tex
\section{Assumption F: Training and Inference Economics}

\textbf{Assumption F: Canonical code reduces required training data, active inference compute, failed repairs, and review burden enough to lower cost per verified correct change.}

Training compute for dense transformers is commonly approximated as
\begin{equation}
  C_{\text{train}} \approx 6ND,
\end{equation}
where $N$ is parameter count and $D$ is training tokens~\cite{kaplanScaling,hoffmannChinchilla}. The canonical program may attack both terms, but that remains a scaling-curve hypothesis. It should be stated as paired tokens-to-target-accuracy, not as a naked training-speed claim. At fixed hardware tokens/s, tokens-to-target reduction is also training-time reduction; equivalently, physical tokens/s become raw-corpus-equivalent tokens/s multiplied by $R_{\text{train}}$. A defensible central target is 30$\times$--150$\times$ lower tokens-to-target capability under successful paired scaling curves, with additional savings from lower active inference, fewer reasoning/tool tokens, fewer failed repair loops, and lower review burden.

For inference, the dense-active speed claim is deliberately narrower than the verified-change claim. If a 100B--120B dense canonical specialist reaches the same accepted-change quality as a 1T-class dense-active baseline on supported canonical work, then first-order active-parameter accounting implies roughly 8.3$\times$--10$\times$ more same-hardware aggregate token/sec and roughly 88\%--90\% lower dense-active infrastructure cost/token before memory, batching, context, and utilization corrections. That claim does not apply unchanged to sparse mixture-of-experts systems: a 1T-total MoE with 32B active parameters may already be cheap per token, so the canonical MoE claim must be measured through router entropy, expert specialization, and end-to-end cost per verified change rather than nominal total parameter count.

The all-in economics of canonical software intelligence are:
\begin{equation}
\begin{split}
 C_{\text{correct}} ={}&
 A_{\text{foundry}} + C_{\text{train}} + C_{\text{infer}} +
 C_{\text{tools}} \\
&+ C_{\text{verify}} + C_{\text{review}} +
 C_{\text{failed}} + C_{\text{maint}}.
\end{split}
\end{equation}
The foundry is a capital expenditure on the training substrate. Its cost must be amortized across repeated model runs, served tokens, downstream agent tasks, security repairs, and reusable certified cells. Active inference compute includes context tokens, reasoning tokens, action/tool tokens, verification-log interpretation, and repair attempts. The strongest business metric is therefore cost per verified correct change, not raw token price. The central specialist scenario is a matched-quality conditional, not a proven empirical result: if the 100B--120B model fails to match broad-model accepted-change quality on supported canonical tasks, then the dense-active speed and cost/token scenario does not activate.

The simplest break-even condition is:
\begin{equation}
k^*=\frac{C_{\mathrm{foundry}}}{C_{\mathrm{raw/change}}-C_{\mathrm{canon/change}}}.
\end{equation}
Here $k^*$ is the number of accepted changes required to repay the foundry cost under a fixed task distribution. If the canonical path does not reduce same-quality change cost, or if verification/governance cost erases the savings, the economic claim fails even when source and action-space metrics improve.

\begin{table*}[t]
\caption{Cost Components Hidden by Raw Token Pricing}
\label{tab:costcomponents}
\centering
\footnotesize
\begin{tabularx}{\textwidth}{p{0.20\textwidth}X X}
\toprule
Component & Raw repository burden & Canonical compression route \\
\midrule
Training compute & Long-tail languages, frameworks, layouts, generated drift, and weak labels. & Paired scaling curves on canonical training objects and behavior graphs. \\
Active inference compute & Long context reads, architecture rediscovery, exploratory tool use. & Reasoning digests, file grammar, generated-zone manifests, proof-carrying change objects. \\
Tool calls & Search, test discovery, command guessing, dependency archaeology. & Proof lanes, repository policy, dependency closure, standard receipts. \\
Verification cost & Ad hoc commands, flaky gates, unclear migration safety. & Deterministic proof lanes, migration replay, rollback receipts, policy checks. \\
Review burden & Reviewers reconstruct intent and proof from a raw diff. & Intent, typed diff, security delta, test delta, receipts, and rollback plan in one object. \\
Failed repairs & Invalid edits, generated-file modifications, unsafe migrations, repeated test failures. & Constrained edit grammar, semantic patch cells, negative-path curriculum. \\
\bottomrule
\end{tabularx}
\end{table*}

\begin{table*}[t]
\caption{Canonical Model Training Ladder}
\label{tab:ladder}
\centering
\footnotesize
\begin{tabularx}{\textwidth}{p{0.16\textwidth}X X}
\toprule
Stage & Model/data configuration & Purpose \\
\midrule
1B--3B scout & Canonical syntax, file grammar, naming, schema, and build command data. & Verify that the standard is learnable under small-model budgets. \\
7B--14B repairer & Messy-to-canonical traces, generated-boundary repairs, migrations, dependency updates. & Measure how much porting logic fits in small specialists. \\
32B--70B system model & Full canonical repositories with long-context tasks, UI/product changes, DB changes, and security fixes. & Establish capability curve before 100B-class runs. \\
100B--120B central model & Full canonical corpus, behavior fixtures, profile routing, and long-horizon planning data. & Central target for matching broad dense-active coding models on canonical work. \\
Mixture-of-experts canonical frontier & Experts for product surface, service/core, durable data and migration, contracts/generated artifacts, security/policy, infra/release, verifier/repair, and porting. & Reduce active compute and improve router precision. \\
\bottomrule
\end{tabularx}
\end{table*}

\textbf{Falsification:} Assumption F fails if empirical scaling curves show no reduction in tokens to target canonical accuracy, if active inference and reasoning tokens do not fall on paired change tasks, if the 8.3$\times$--10$\times$ dense-active token/sec scenario disappears after memory, batching, context, utilization, verification, or quality corrections, if a canonical specialist needs more than 50\% of broad dense-active parameters to match broad-model accuracy on held-out canonical tasks, or if foundry cost does not amortize within 3$\times$ the initial investment.


%% file: sections/11_assumption_g_role_moe.tex
\section{Assumption G: Canonical Role Mixture-of-Experts}

\textbf{Assumption G: Canonical code gives mixture-of-experts models cleaner expert domains, reducing router entropy and improving expert utilization.}

Kimi K2 shows why dense and mixture-of-experts baselines must be separated: it reports 1T total parameters but only about 32B active parameters per token~\cite{kimiK2,kimiDocs}. A dense 100B canonical model is not automatically faster than a 32B-active mixture-of-experts model on raw per-token compute. The canonical opportunity for mixture-of-experts is different.

The human-code mixture-of-experts model must route across many languages, frameworks, and repository idioms. A canonical-code mixture-of-experts model routes across fewer, cleaner domains: product surface, service/core, durable data and migration, contracts/generated artifacts, security policy, dependency governance, verification, repair, and planning. Experts specialize around canonical software roles rather than arbitrary language/framework combinations.

This matters because routing uncertainty is another form of human-code entropy. In a raw repository, a billing change might require a controller, browser form, secret, CI permission, SQL migration, queue worker, and webhook verifier, each following local conventions. In the canonical substrate, those concerns have named roles, owned files, generated boundaries, and proof lanes. The router can learn the role graph directly: billing cell, policy cell, SQL migration expert, UI resource-state expert, verifier/repair expert. The model spends less capacity deciding what kind of software world it is inside.

The canonical mixture-of-experts hypothesis is therefore not merely ``use sparse models.'' It is that canonicalization gives sparse models better expert boundaries. A role expert can specialize in migration invariants rather than every dialect of migrations; a security expert can specialize in policy edges and negative cases rather than every framework's middleware idiom; a verifier expert can specialize in proof receipts and failed lanes rather than repository-specific CI folklore.

The metrics are explicit. Router entropy should fall for a fixed task distribution:
\begin{equation}
H_{\mathrm{router}}(x)=-\sum_e p(e\mid x)\log p(e\mid x).
\end{equation}
Success also requires expert load balance, low dropped-token or overflow rate, specialization purity by software role, low cross-expert repair churn, and equal or better task success at fixed active parameters. A canonical role mixture-of-experts should be compared against dense models and broad mixture-of-experts baselines by accepted-patch rate and cost per verified correct change, not by nominal total parameter count.

\begin{table}[t]
\caption{Canonical Role Mixture-of-Experts Layout}
\label{tab:rolemoe}
\centering
\footnotesize
\begin{tabularx}{\columnwidth}{p{0.26\columnwidth}X}
\toprule
Expert family & Canonical role \\
\midrule
Product surface & UI state, accessibility, visual behavior, and typed user-facing flows. \\
Service/core & Domain logic, APIs, adapters, and service invariants. \\
Durable data/migration & Durable truth, schema changes, migration safety, and rollback. \\
Contracts/generated & Schemas, generated clients, fixtures, adapters, and docs. \\
Security/policy & Authentication, authorization, input boundaries, secrets, dependency risk. \\
Infra/release & Deployment envelopes, release policy, observability, and operational permissions. \\
Verifier/repair & Validation lanes, failed traces, blocked changes. \\
Porter & Messy-to-canonical translation and disposition assignment. \\
\bottomrule
\end{tabularx}
\end{table}

\textbf{Falsification:} Assumption G fails if router entropy, expert utilization, or task success do not improve versus language/framework-oriented mixture-of-experts routing.


%% file: sections/12_assumption_h_evaluation.tex
\section{Assumption H: Evaluation and Falsification}
\label{sec:eval}

\textbf{Assumption H: The canonical substrate reduces invalid actions and end-to-end change cost, not merely produces cleaner-looking code.} The evaluation target is not HumanEval-style function synthesis. The target is whether a model operating inside the canonical universe can implement, modify, test, migrate, secure, and deploy ported software at lower cost than a broad model operating inside the human-chaotic universe. SWE-bench Verified is a useful precedent~\cite{swebench,swebenchVerified}, but static issue benchmarks are vulnerable to weak tests, contamination, and benchmark aging~\cite{utboost,sweRebench,openaiSwebenchVerifiedCritique}. This paper's claim needs paired raw/canonical tasks from the same software lineage.

The benchmark object is the tuple:
\begin{quote}
\footnotesize
(\texttt{raw\_repo}, \texttt{canonical\_repo}, \texttt{issue}, \texttt{behavior\_contract},\\
\texttt{preservation\_tier}, \texttt{proof\_lanes}, \texttt{hidden\_tests},\\
\texttt{reviewer\_rubric}, \texttt{cost\_ledger})
\end{quote}

The proof program has three gates. \textbf{Gate 1} is the same-model substrate test: run the same broad frontier model on raw and canonical versions of the same issue lineage. This is the first falsification point because it isolates substrate value before specialist training. \textbf{Gate 2} trains or adapts a canonical specialist and compares it on canonical tasks against the same broad baseline and the raw/canonical same-model result. \textbf{Gate 3} measures paired scaling curves and verified-change economics: tokens-to-target, model size, serving cost, foundry amortization, and all-in cost per accepted proof-carrying change.

\begin{table*}[t]
\caption{Canonical Port Evaluation Arms}
\label{tab:evalarms}
\centering
\footnotesize
\begin{tabularx}{\textwidth}{p{0.06\textwidth}p{0.24\textwidth}p{0.22\textwidth}X}
\toprule
Arm & Model & Repository form & Purpose \\
\midrule
A & Broad frontier model & Original human repo & Current-world baseline: capability and cost in the messy corpus. \\
B & Broad frontier model & Canonical repo & Measures substrate benefit without a new model. \\
C & Canonical specialist & Canonical repo & Measures model plus substrate effect. \\
D & Canonical specialist + cells & Cellized canonical repo & Measures behavior-cell compression benefit. \\
E & Porter model & Human repo to canonical object & Measures foundry automation throughput. \\
\bottomrule
\end{tabularx}
\end{table*}

The key ablation is direct: \textbf{if the same broad frontier model performs much better on canonical repos than on the original raw repos, before canonical specialist training, then substrate compression is real.} That proof does not require waiting for a 100B-class specialist. The paired task must measure accepted patch rate and the full trajectory: files opened, context tokens, reasoning tokens, tool calls, invalid edits, generated-zone violations, failed tests, failed migrations, wall time, reviewer comments, cost ledger entries, and rollback/proof receipts.

\begin{table*}[t]
\caption{Killer Ablation: Raw Repo Versus Canonical Repo with the Same Model}
\label{tab:killerablation}
\centering
\footnotesize
\begin{tabularx}{\textwidth}{p{0.18\textwidth}X X}
\toprule
Metric & Raw repository expectation & Canonical repository target \\
\midrule
Files opened & Broad search across local architecture. & Fewer predictable file reads from canonical path grammar. \\
Tool calls & Discover build, tests, migrations, policy, dependencies. & Invoke known proof lanes and typed patch tools. \\
Reasoning tokens & Rediscover architecture, ownership, generated zones, repair strategy. & Use reasoning digest, cell contracts, and proof obligations. \\
Invalid edits & Generated-file edits, wrong layer, unsafe migration, policy bypass. & Rejected by constrained edit grammar before proof lanes. \\
Failed verification & Flaky or misselected commands and repeated repair loops. & Standard receipts and classified failures. \\
Accepted patch rate & Baseline for current-world agents. & Higher acceptance at equal or lower total cost. \\
\bottomrule
\end{tabularx}
\end{table*}

\begin{table*}[t]
\caption{Contamination-Aware Benchmark Tiers}
\label{tab:benchmarktiers}
\centering
\footnotesize
\begin{tabularx}{\textwidth}{p{0.18\textwidth}X X}
\toprule
Tier & Use & Risk controlled \\
\midrule
Public paired & Reproducible raw/canonical ablations and open leaderboard tasks. & Transparent but contamination-prone; not sufficient for frontier claims. \\
Private held-out & Partner repositories, undisclosed issues, hidden tests, and blinded review. & Reduces training contamination and patch memorization. \\
Live / post-cutoff & New issues evaluated before public disclosure and paired canonicalization. & Tests current repository reasoning under fresh task distributions. \\
Generated stress tasks & SWE-smith-style generated task families and preservation traps~\cite{sweSmith}. & Scales coverage and tests false-green, generated-zone, and compatibility-ghost failures. \\
\bottomrule
\end{tabularx}
\end{table*}

\begin{table*}[t]
\caption{Canonical Standard Ablation Ladder}
\label{tab:ablation}
\centering
\footnotesize
\begin{tabularx}{\textwidth}{p{0.20\textwidth}X X}
\toprule
Ablation & What changes & What it proves \\
\midrule
Naming only & Controlled vocabulary and role names; no layout or dependency changes. & Separates lexical entropy from structural entropy. \\
Path grammar only & Canonical folders, ownership, test locations, generated/source paths. & Measures repository-navigation and edit-location reduction. \\
Generated boundaries only & Contracts generate clients/adapters/fixtures; hand edits forbidden. & Measures duplicate-truth and invalid generated-edit reduction. \\
Repository policy only & Branch protection, code owners, required checks, workflow permissions, secrets, release gates. & Measures hidden-state and CI/action-space reduction. \\
Dependency governance only & Pinned, owned, scored dependencies and upgrade lanes. & Measures supply-chain and repair-loop reduction. \\
Proof lanes only & Standard build/test/migration/security commands and acceptance artifacts. & Measures failed-loop reduction without full language/profile collapse. \\
Behavior cells only & Top 50 cells with generated expansion. & Measures source-token and action-space reduction from cells alone. \\
Full standard & Language roles, file grammar, naming, generated boundaries, repository policy, dependencies, data truth, cells, validation. & Measures total canonical substrate effect. \\
\bottomrule
\end{tabularx}
\end{table*}

\begin{table*}[t]
\caption{Core Assumptions, Strength Assessments, and Falsification Criteria}
\label{tab:assumptions}
\centering
\footnotesize
\begin{tabularx}{\textwidth}{p{0.03\textwidth}p{0.25\textwidth}p{0.12\textwidth}X}
\toprule
ID & Assumption & Strength & Falsified if \\
\midrule
A & One governed way can cover each supported product-software concern. & Strong & Primary profile covers less than 60\% of economically valuable product/web/backend/data tasks. \\
B & Full-corpus foundry can assign governed dispositions and amortize cost. & Hard & Large fractions remain undispositioned or foundry cost exceeds downstream savings. \\
C & Behavior can be preserved or dispositioned with graded confidence. & Hardest & Human review rejects core ports or security/migration regressions rise. \\
D & Behavior cells absorb high-reuse routine product code. & Very plausible & Top 500--2,000 cells cover less than 50\% of product-app behavior. \\
E & Canonicalization reduces representation entropy and legal action-space. & Central claim & Paired corpora show no entropy, path, AST, perplexity, or action-branching reduction. \\
F & Canonical data reduces required model size and training tokens. & Hypothesis & Scaling curves show no lower token or parameter need for target canonical accuracy. \\
G & Canonical mixture-of-experts routes better by software role than by language accident. & Plausible & Router entropy, utilization, and task success do not improve. \\
H & Canonical substrate lowers cost per verified correct change. & Business claim & End-to-end change cost is not at least 3$\times$ lower after amortization. \\
I & Canonical substrate compresses reasoning, tool, planning, and retry tokens. & Moonshot claim & Same-model canonical tasks do not reduce files opened, tool calls, reasoning tokens, invalid edits, failed tests, wall time, or reviewer burden. \\
\bottomrule
\end{tabularx}
\end{table*}

Major risks are first-class evaluation targets and are expanded in Section~\ref{sec:threatmodel}. \textbf{Licensing and provenance} must travel with every source artifact; The Stack v2 and the BigCode Governance Card both emphasize governance, provenance, privacy, and consent obligations for code data~\cite{theStackV2,bigcodeGovernanceCard}. Software Heritage's 2025 activity report is a reminder that the public software archive is enormous and must be treated as infrastructure, not scrapeable exhaust~\cite{softwareHeritage2025}. \textbf{Compression-factor overlap} is handled by paired-corpus measurement; the paper never multiplies language collapse, framework collapse, cell collapse, action-space collapse, reasoning-token collapse, and repair-path collapse naively. \textbf{Representational compression is not automatically model compression}; that requires scaling curves.

\textbf{Falsification:} Assumption H fails if cost per verified correct change is not at least 3$\times$ lower after including foundry, verification, serving, reasoning, failed-loop, review, and maintenance cost amortization.


%% file: sections/12a_cost_ledger_field_trial.tex
\section{Cost Ledger and Field Trial Design}
\label{sec:cost-ledger}

The paper should be judged by one primary endpoint: \textbf{amortized cost per verified correct change}. A result that reduces source tokens but increases proof failures is a loss. A result that improves benchmark pass rate but increases reviewer burden is incomplete. A result that lowers token cost while increasing incident remediation is not a win. The ledger must include source, context, reasoning, and action tokens; serving infrastructure; verification; security and provenance work; tool runtime; wall time; human review; failed attempts; rollback work; downstream defect cost where measurable; and foundry amortization.

\begin{equation}
R_{\mathrm{change}}=\frac{C^{\mathrm{raw}}_{\Delta}+A^{\mathrm{raw}}_{\mathrm{foundry}}/N}{C^{\mathrm{canon}}_{\Delta}+A^{\mathrm{canon}}_{\mathrm{foundry}}/N},
\end{equation}
where $A_{\mathrm{foundry}}$ is foundry construction and porting cost amortized across $N$ future changes. The aggressive case is impossible if the foundry never amortizes; the conservative case can be true even while the foundry is still expensive if repeated verified changes become much cheaper.

\begin{table*}[t]
\caption{Minimum Ledger for a Defensible Field Trial}
\label{tab:cost-ledger}
\centering
\footnotesize
\begin{tabularx}{\textwidth}{p{0.20\textwidth}X X}
\toprule
Ledger item & Raw repository measurement & Canonical repository measurement \\
\midrule
Discovery & Files opened, grep/search calls, dependency reads, architecture notes. & Cell lookups, profile docs read, reasoning digest reads. \\
Context/pre-fill & Input tokens, cache hit rate, long-context latency. & Reduced context tokens, generated-zone manifests, cached profile/cell context. \\
Planning/reasoning & Planning tokens, hidden reasoning budget where measurable, self-repair tokens, tool-selection deliberation. & Typed change recognition, bounded proof-plan tokens, no-reason lanes. \\
Generation/action & Output tokens, patch size, tool calls, action tokens. & Typed operation, generated diff, proof object, rollback object. \\
Serving & Dense-active or MoE cost/token, token/sec, memory, batching, utilization. & Matched-quality canonical specialist serving ledger plus verification overhead. \\
Verification & Test commands, failed lanes, log interpretation, flaky reruns. & Standard proof-lane receipts, classified failures, deterministic replay. \\
Security/provenance & SBOM, license/provenance review, secret scan, policy checks. & Carried metadata, transformation receipts, governed policy gates. \\
Review & Reviewer comments, review minutes, rework cycles. & Intent audit, proof receipt audit, residual novelty audit. \\
Reliability & Escaped defects, incident tickets, rollback time. & Runtime invariant violations, negative-memory updates, rejected cells. \\
Amortization & None or local scripts only. & Foundry cost spread over future changes, repos, profiles, and cells. \\
\bottomrule
\end{tabularx}
\end{table*}

The first field trial should avoid the common benchmark trap of measuring only task success. It should choose 20--50 production-like repositories with known issue lineage and hidden reviewer rubrics, port each into the canonical substrate, and run arms A--D from Table~\ref{tab:evalarms}. The decisive result is not that the canonical specialist wins. The decisive result is that the same broad model in arm B becomes cheaper and more reliable than arm A. That isolates substrate value from model value. Then arm C tests specialist training, and arm D tests behavior-cell amortization.

The paper should publish every failure. Failed ports reveal weak behavior oracles. Compatibility ghosts reveal hidden contracts. Non-preserved behavior reveals product decisions. Negative examples become training material. The foundry should not hide these cases; it should classify them. A canonical corpus with honest rejected dispositions is more valuable than a clean-looking corpus that silently drops behavior.


%% file: sections/13_risks_and_threat_model.tex
\section{Risks and Threat Model}
\label{sec:threatmodel}

The canonical substrate is not risk-free. It deliberately concentrates software structure, so its failure modes must be named as design constraints rather than left as caveats.

\begin{table*}[t]
\caption{Canonical Substrate Threat Model}
\label{tab:threatmodel}
\centering
\footnotesize
\begin{tabularx}{\textwidth}{p{0.20\textwidth}X X}
\toprule
Risk & Failure mode & Required mitigation \\
\midrule
Behavior loss & Port drops hidden edge behavior, migration side effects, timing assumptions, or compatibility ghosts. & Declared contracts, preservation tiers, differential replay, accepted-incompatibility manifests, and human escalation. \\
Weak oracles & Tests pass while behavior diverges. & Treat tests as evidence only; add fuzzing, property tests, mutation tests, generated tests, hidden review, and trace replay. \\
Licensing/provenance & Canonical artifact loses attribution, opt-out state, source lineage, or derivative-risk status. & SPDX/CycloneDX metadata, source hashes, transformation traces, removal propagation, and policy gates. \\
Canonical monoculture & One flawed auth, migration, billing, or policy cell creates correlated failure at scale. & One governed interface with independent implementations for critical cells, conformance suites, canaries, rollback, and cell advisories. \\
Useful-diversity loss & Profile excludes a domain-specific implementation choice that carries performance, safety, accessibility, regulatory, or ecosystem value. & Secondary profiles, governed primitives, explicit exception process, and profile lifecycle evidence. \\
Benchmark contamination & Public paired tasks leak into training or static tests age out. & Public, private held-out, and live/post-cutoff benchmark tiers. \\
Foundry cost & Porting, proof, review, and governance costs exceed downstream savings. & Break-even accounting, staged dispositions, high-reuse prioritization, and amortization receipts. \\
Governance capture & Canonical rules favor one vendor, stack, or implementation without evidence. & Versioned public specs, independent conformance suites, audit logs, and appeal paths. \\
Generated-code opacity & Source becomes generated but reviewers cannot understand obligations or failures. & Proof-carrying change objects, generated-zone manifests, readable diffs, and reviewer rubrics. \\
Scaling-curve uncertainty & Representation compression does not translate to training-token or model-size reduction. & Paired raw/canonical scaling curves and fixed-target accepted-change benchmarks. \\
\bottomrule
\end{tabularx}
\end{table*}

The central controlled-diversity rule is: canonicalize behavior interfaces, proof receipts, and edit grammars; diversify critical implementations and evaluators where common-mode failure would be catastrophic. N-version programming is a warning as much as a precedent: diversity is useful only when failures are sufficiently independent~\cite{nVersionProgramming}. Unlimited local dialects are not useful diversity. Independently validated lowerings, fuzzers, policy engines, monitors, and rollback paths are.


%% file: sections/13_compression_frontier.tex
\section{Compression Frontier Summary}
\label{sec:compression-frontier}

The full 30-domain compression catalog is supplemental source in \texttt{compression\_catalog.tex}. The main paper keeps only the proof spine: every domain is a prior or target until paired measurement promotes it. The catalog is useful for search-space accounting, but it is not the evidence base by itself.

\begin{table*}[t]
\caption{Main-Paper Compression Frontier Summary}
\label{tab:frontier-summary}
\centering
\footnotesize
\begin{tabularx}{\textwidth}{p{0.18\textwidth}p{0.18\textwidth}X X}
\toprule
Layer & Current status & Main mechanism & Required measurement \\
\midrule
Corpus hygiene & Partly measured in prior corpora & Deduplication, generated/vendor/secret/malware filtering, provenance assignment. & Raw/canonical token counts, removal receipts, license disposition rates. \\
Canonical profile & Near-term target & Fixed roles, path grammar, generated zones, proof lanes, dependency law, repository policy. & Same-model raw/canonical files opened, invalid edits, proof-lane failures, accepted patch rate. \\
Contract-first generation & Central hypothesis & Schemas generate clients, fixtures, validators, docs, migrations, and adapters. & Duplicate-truth removal, generated-edit violations, schema drift rate, migration failures. \\
Behavior cells & Central hypothesis & Auth, resource lifecycle, search, billing, uploads, jobs, webhooks, audit, observability as certified cells. & Behavior-cell census by accepted changes, AST nodes, traces, issues, review burden, and security paths. \\
Semantic patch cells & Central hypothesis & Typed edits such as add field, split table, rotate secret, add idempotency key, and add permission edge. & Legal action count, invalid edit count, proof receipt completeness, repair-loop reduction. \\
Runtime and negative memory & Moonshot & Incidents, reverted patches, production traces, and rejected plans become invariants and forbidden paths. & Incident recurrence, compatibility-ghost capture, negative-test reuse, post-deployment regression rate. \\
\bottomrule
\end{tabularx}
\end{table*}

The frontier should be read as an accounting program. Literal source-token reduction is expected to be modest compared with legal-action and reasoning-space reduction. Training-token reduction is the hardest claim and remains conditional on paired scaling curves. Cost per verified correct change is the final denominator because it includes foundry, verification, serving, reasoning, failed-loop, review, and maintenance cost.


%% file: sections/14_future_vision.tex
\section{Future Vision: Minimum Viable Novelty}
\label{sec:futurevision}

The long-term prize is not a smaller corpus of cleaner code. It is a post-source-code substrate in which routine software is compiled from compact intent, behavior cells, invariants, policies, proofs, provenance, runtime evidence, and reusable reasoning. Source code remains useful, inspectable, and deployable, but it becomes one generated projection of a deeper object.

This is the extreme form of \mfdl{}: the system should contain only \textbf{minimum viable novelty}. Humans still decide product goals, legal obligations, risk tolerance, taste, and genuinely new domain behavior. But all routine coding decisions---language, framework, folder, generated boundary, migration idiom, dependency wrapper, proof lane, review evidence, rollback form, and repair strategy---move out of human discretion and into governed substrate law.

\begin{table*}[t]
\caption{Three Build-Out Stages for the Canonical Substrate}
\label{tab:futurestages}
\centering
\scriptsize
\begin{tabularx}{\textwidth}{p{0.13\textwidth}X X X}
\toprule
Stage & Substrate built & Main compression denominator & Target range \\
\midrule
1. Canonical repo & Primary/secondary profiles, file grammar, generated zones, proof lanes, repository policy. & Training tokens, active context, tool discovery, invalid edits. & \conrange{3$\times$--10$\times$ training; 2$\times$--8$\times$ inference/reasoning/tool; 2$\times$--5$\times$ verified-change cost.} \\
2. Correct-change substrate & Behavior cells, semantic patch cells, proof-carrying change objects, reasoning digests, negative corpus. & Action space, repair loops, review burden, repeated reasoning. & 30$\times$--150$\times$ training; 10$\times$--100$\times$ inference/reasoning/tool; 10$\times$--50$\times$ verified-change cost. \\
3. Behavior genome & Behavior IR, software genome, runtime-derived invariants, agent-native OS, verification markets, evolutionary foundry. & Routine-domain training, action/reasoning/search space, institutional memory, non-novel implementation. & \aggrange{150$\times$--1,000$\times$ routine-domain training; 100$\times$--10,000$\times$ inference/action/reasoning; 50$\times$--1,000$\times$ verified-change cost.} \\
\bottomrule
\end{tabularx}
\end{table*}

The defensible path starts with facts already visible in the field. Public software archives are massive and structured; Software Heritage reported over 27 billion unique source files from 421 million projects in its 2025 activity report~\cite{softwareHeritage2025}. Raw code is repetitive and predictable~\cite{hindleNaturalness}; large clone studies and notebook studies show heavy duplication~\cite{dejavu2017,jupyterClones2020}. Product-line engineering has long treated software families as managed commonality plus variability~\cite{clementsProductLines}. LLVM, MLIR, e-graphs, OpenAPI, Protocol Buffers, and Coccinelle are all partial precedents for moving from raw text toward intermediate representations, generated projections, equivalence classes, contracts, and semantic transformations~\cite{lattnerLLVM,lattnerMLIR,eggEgraphs,openapi31,protobufOverview,coccinelle}. Proof-carrying code, CompCert, seL4, and SLSA show that evidence and provenance can become first-class artifacts rather than after-the-fact paperwork~\cite{proofCarryingCode,compcertManual,sel4Proofs,slsa}.

\textbf{Software genome.} The foundry should mine raw repositories, issues, tests, traces, incidents, dependency histories, and repair patches into a global atlas of behavior families: authentication, tenant boundaries, resource lifecycle, billing, uploads, jobs, webhooks, audit, observability, migrations, notifications, permissions, retries, cache invalidation, and product workflows. The output is not a snippet library. It is a behavior gene bank: canonical intent, variants, anti-examples, contracts, tests, proof templates, repair memories, provenance, legal transformations, and generated implementations. DreamCoder-style library learning supports the general idea that learned abstractions can shorten future program search~\cite{dreamcoder}; the canonical foundry applies that logic to production software behavior.

\textbf{Behavior IR.} The substrate needs an application-level intermediate representation above source code. A behavior IR represents entities, state machines, policies, permissions, effects, external contracts, migration semantics, observability, runtime SLOs, and proof obligations. It lowers into governed service code, typed product surfaces, durable-data operations, UI, docs, tests, dashboards, deployment manifests, and formal/specification artifacts. This is the product-software analogue of compiler IR: agents edit semantic deltas, not arbitrary file trees.

\textbf{Typed change algebra.} Routine changes should become typed operations with known proof obligations:
\begin{table}[t]
\caption{Typed Change Algebra Examples}
\label{tab:typed-change-algebra}
\centering
\scriptsize
\begin{tabularx}{\columnwidth}{p{0.42\columnwidth}X}
\toprule
Operation & Parameters \\
\midrule
\texttt{AddField} & \texttt{User, timezone, nullable=false, backfill=UTC} \\
\texttt{AddPermissionEdge} & \texttt{actor=Manager, action=ApproveInvoice} \\
\texttt{SplitTable} & \texttt{source=Events, target=AuditEvents} \\
\texttt{AddIdempotencyKey} & \texttt{endpoint=WebhookDispatch} \\
\texttt{RotateSecret} & \texttt{provider=Stripe} \\
\bottomrule
\end{tabularx}
\end{table}
Each operation expands into schema diff, migration plan, generated client updates, UI state changes, security review, observability deltas, tests, rollback, and receipts.

\textbf{Reasoning compiler.} The most important future gain is cognitive amortization. Successful and failed trajectories should compile into reusable plans: issue type to affected cells, repository profile to legal actions, failure signature to repair strategy, migration class to proof obligations, dependency update to compatibility checks. Current reasoning and acting methods show why intermediate reasoning helps~\cite{chainOfThought,selfConsistency,react}; the canonical future is to spend that reasoning once, store it as substrate law, and execute known work through no-reason or bounded-reason lanes.

\textbf{Proof-carrying changes.} Future agents should not submit raw diffs. They should submit typed change objects containing intent, affected behavior cells, semantic patch, generated source diff, schema diff, migration diff, tests, security delta, proof obligations, receipts, rollback plan, and provenance/license metadata. Human review shifts from reconstructing intent from text to auditing changed obligations, assumptions, and residual novelty.

\textbf{Evolutionary foundry.} Once behavior cells have evaluators, cells can improve continuously. Agents propose variants; tests, proofs, fuzzers, benchmarks, security scanners, and runtime monitors select survivors. AlphaDev and FunSearch are narrow but important evidence that evaluator-guided program search can discover useful algorithms beyond ordinary human implementation~\cite{alphadev,funsearch}. The canonical version extends the loop to product behavior: better retries, indexes, migrations, policy encodings, UI state machines, rollback paths, and proof envelopes.

\textbf{Runtime and negative memory.} The substrate should learn from production traces, incidents, rejected patches, reverted commits, review objections, failed migrations, security bugs, flaky tests, and rollbacks. Runtime behavior reveals implicit compatibility contracts; negative examples become forbidden plans, red-team tests, proof obligations, and repair memories. This is how the system avoids destroying hidden behavior while still compressing historical accident.

\textbf{Verification markets and agent-native software OS.} A mature substrate needs independent evaluators: proof-lane providers, fuzzing services, conformance suites, cell auditors, provenance validators, and runtime monitors competing on evidence quality. The agent-facing operating system is then not a terminal plus repository checkout. It is a governed environment exposing typed changes, behavior cells, proof lanes, policy decisions, provenance receipts, runtime traces, and negative memory as native system calls.

The extreme endpoint is not zero code and not zero engineering judgment. It is \textbf{no accidental software}: no accidental architecture, no accidental dependency choice, no accidental CI, no accidental migration ritual, no accidental security policy, no accidental review burden, and no repeated reasoning where the substrate has already learned the law. The human and agent frontier moves to irreducible novelty.


%% file: sections/14a_first_six_experiments.tex
\section{The First Six Experiments That Decide the Thesis}
\label{sec:first-six-experiments}

The paper becomes stronger if it stops asking readers to believe a grand end state and instead offers a near-term kill chain. The first decisive program should be small enough to run before a full foundry exists and strong enough that a negative result hurts.

\begin{table*}[t]
\caption{Six Experiments That Convert the Thesis into Evidence}
\label{tab:first-six-experiments}
\centering
\footnotesize
\renewcommand{\arraystretch}{1.14}
\begin{tabularx}{\textwidth}{p{0.18\textwidth}X X}
\toprule
Experiment & Pass condition & Fails the thesis if \\
\midrule
Same-model substrate ablation & A broad frontier or strong open model solves paired canonical tasks with materially fewer files, tokens, tool calls, failed loops, and reviewer comments than raw tasks. & Canonical repositories look cleaner but do not reduce search work or accepted-change cost. \\
Training tokens to accepted-change target & Same model family reaches fixed accepted-change quality with fewer canonical tokens than raw tokens, with confidence intervals. & Representation compression does not become sample-efficiency gain. \\
Reasoning/tool/retry reduction & Paired canonical tasks show fewer planning tokens, hidden reasoning tokens where measurable, tool calls, invalid edits, proof failures, validation reruns, and retries. & The model spends the same search budget despite the canonical substrate. \\
Foundry behavior preservation & Ports survive hidden tests, fuzzing, mutation tests, replay, migration checks, security negatives, and human adjudication. & Canonicalization passes easy tests while losing compatibility ghosts or security behavior. \\
Cost per verified correct change & Porting, proof, governance, serving, verification, review, and maintenance costs are repaid by repeated accepted-change savings. & Foundry cost or verification overhead erases downstream economics. \\
Specialist versus broad model & A canonical specialist matches or beats a broad model on supported canonical work at lower all-in cost. & Supported-domain specialization fails to improve accepted-change quality, speed, or cost after corrections. \\
\bottomrule
\end{tabularx}
\end{table*}

These experiments also protect the paper from its own ambition. The 150$\times$--1,000$\times$ story is not the starting claim; it is the long-horizon envelope after cells, proof lanes, negative memory, and behavior IR mature. The starting claim is harsher and cleaner: if same-model raw/canonical ablation does not produce obvious cost and search-work savings, the training-token moonshot should not be believed.


%% file: sections/14b_frontier_extensions.tex
\section{Frontier Extensions That Make the Thesis Harder to Ignore}
\label{sec:level-up}

The current paper should be read as a substrate thesis plus a research agenda. The strongest next ideas are not cosmetic; they turn the claim from a compression manifesto into a defensible experimental program.

\begin{table*}[t]
\caption{Highest-Leverage Additions for a Stronger Canonical-Code Program}
\label{tab:levelup}
\centering
\footnotesize
\begin{tabularx}{\textwidth}{p{0.20\textwidth}X X}
\toprule
\textbf{Addition} & \textbf{Why it matters} & \textbf{Concrete artifact} \\
\midrule
Behavior-cell coverage census & Converts moonshot language into a measured market map: what fraction of product software is resource lifecycle, auth, policy, workflow, billing, notification, observability, integration, or migration routine? & A labeled corpus with per-file and per-change coverage by cell family, plus residual novelty estimates. \\
Canonicalization loss function & Prevents the foundry from optimizing prettiness instead of preserved behavior and lower change cost. & Multi-objective score: behavior preservation, proof strength, source entropy, action entropy, security posture, provenance quality, and amortized cost. \\
Paired repository arena & Makes the decisive claim testable without waiting for full-corpus porting. & Raw/canonical repository pairs, identical issues, hidden tests, proof lanes, reviewer rubrics, and full token/tool/dollar traces. \\
Negative memory bank & Turns incidents, reverted patches, flaky tests, security bugs, and failed migrations into reusable anti-examples. & Versioned forbidden-plan cells and regression generators attached to behavior cells and proof lanes. \\
Renderer pluralism & Keeps the profile ambitious without becoming brittle or cultish. & One behavior IR rendered to the primary product profile and to secondary governed profiles where domain constraints require it. \\
Foundry amortization ledger & Makes the economics credible. & Per-port cost, review cost, proof cost, reuse count, served-token savings, accepted-change savings, and break-even $k^*$. \\
Adversarial equivalence audit & Protects against canonical ports that pass easy tests while losing edge behavior. & Differential replay, fuzzing, mutation testing, symbolic checks where feasible, security review, and human product-owner adjudication. \\
\bottomrule
\end{tabularx}
\end{table*}

The most important addition is the behavior-cell coverage census. If 70--90\% of routine product-software changes fall into reusable cells and semantic patch cells, the aggressive thesis becomes plausible. If coverage stalls at 20--30\%, canonical code may still be valuable, but the paper becomes an engineering-efficiency paper rather than a training-substrate revolution. The second most important addition is the paired repository arena, because it directly tests the economic question: can the same or smaller model produce accepted proof-carrying changes with fewer total tokens, tool calls, failed loops, and reviewer interventions?

Four frontier extensions should remain explicitly future work, not current evidence. \textbf{Behavior genomes} would mine raw repositories, traces, incidents, and repairs into a global atlas of behavior families with variants, invariants, tests, proofs, provenance, and generated projections. \textbf{Runtime-derived invariants} would turn production traces, rollbacks, failed migrations, and reverted patches into negative memory and proof obligations. \textbf{Verification markets} would let independent proof-lane providers, fuzzers, conformance suites, cell auditors, provenance validators, and runtime monitors compete on evidence quality. \textbf{Agent-native operating systems} would expose typed changes, behavior cells, proof lanes, policy decisions, provenance receipts, runtime traces, and negative memory as native system calls rather than leaving agents inside an unstructured terminal and repository checkout.

A mature canonical foundry should therefore publish two ledgers. The \emph{corpus ledger} reports what was ingested, rejected, ported, preserved, proven, license-cleared, or quarantined. The \emph{change ledger} reports every accepted and failed task: context tokens, reasoning tokens, tool calls, invalid actions, proof failures, repair loops, wall-clock, review comments, and dollars. The paper's central numbers should rise or fall with those ledgers.


%% file: sections/14c_no_accident_horizon.tex
\section{The No-Accident Horizon}
\label{sec:no-accident-horizon}

The strongest form of this paper is not that all future software becomes free. That would be false for arbitrary future programs and misleading for reviewers. The stronger and defensible claim is narrower: once accidental representation, repeated architecture, duplicated contracts, routine behavior families, proof-route discovery, generated-surface drift, dependency rituals, migration folklore, and repair loops have been quotiented away, software converges to a residual floor. That floor is the \textbf{No-Accident Horizon}.

\subsection{The Limit Question}

Let $A_t$ be the admissible software evidence available at time $t$: repositories, issues, pull requests, tests, traces, incidents, vulnerabilities, reviews, schemas, deployment histories, rejected patches, documentation, provenance, license metadata, and rollback evidence. Let $M_t=\canonmap(A_t)$ be the best governed canonical memory produced from that evidence. The memory is admissible only if it preserves provenance, weak-oracle labels, rejected dispositions, security constraints, migration law, and behavior evidence.

Let $Y\sim P_t(Y\mid A_t)$ be a future software demand drawn from a declared workload distribution, and let $\oracle_{\tau,H}$ be an acceptance oracle at evidence tier $\tau$ and future horizon $H$. The oracle includes behavior, hidden compatibility, security, migration safety, provenance, review policy, and future adaptability. A foundry $F$ maps $(M_t,Y)$ to a proof-carrying change object or to a refusal. Its all-in accepted-change cost is
\begin{align}
C_F(Y)={}&C_{\mathrm{intent}}+C_{\mathrm{context}}+C_{\mathrm{reason}}+C_{\mathrm{tool}}
        +C_{\mathrm{verify}} \notag\\
&+C_{\mathrm{review}}+C_{\mathrm{risk}}+C_{\mathrm{amortized\ foundry}}
        +C_{\mathrm{defect}} .
\label{eq:nah-cost-ledger}
\end{align}
The raw baseline $C_{\mathrm{raw}}(Y)$ is the same ledger for the original human repository and process under the same oracle. The denominator is therefore cost per verified correct change, not lines of code, source bytes, prompt tokens, or benchmark solve rate.

For a distribution and oracle, define the removable-work fraction as
\begin{equation}
\Lambda_{\mathrm{NA}}(P_t,\oracle_{\tau,H})
=1-\frac{\inf_{F\in\mathcal{F}_{\mathrm{adm}}}\mathrm{E}_{Y\sim P_t}[C_F(Y)]}
{\mathrm{E}_{Y\sim P_t}[C_{\mathrm{raw}}(Y)]}.
\label{eq:no-accident-horizon-definition}
\end{equation}
The corresponding ideal multiplier is
\begin{equation}
M_{\mathrm{NA}}=\frac{1}{1-\Lambda_{\mathrm{NA}}}.
\label{eq:no-accident-multiplier}
\end{equation}
This normalization is severe. A $10\times$ result means 90\% removable work, a $100\times$ result means 99\% removable work, and a $1{,}000\times$ result means 99.9\% removable work.

\subsection{No Universal Software Inverse}

\textbf{No universal software inverse.} For arbitrary computable programs or adversarial future software demands, no computable foundry has a positive guaranteed No-Accident Horizon. In the distribution-free case,
\begin{equation}
\inf_{P}\,\Lambda_{\mathrm{NA}}(P,\oracle_{\tau,H})=0.
\label{eq:nah-minimax-zero}
\end{equation}

\noindent\emph{Proof sketch.} If a system could always construct, prove, or correctly reject every arbitrary future behavior at finite bounded residual cost, it could decide nontrivial semantic properties of arbitrary programs by encoding those properties as software intents. That contradicts the Turing--Rice boundary~\cite{turing1936computable,rice1953classes,riceTheoremMIT}. If the future distribution is unconstrained, no-free-lunch reasoning gives the learning-theoretic version of the same warning: an optimizer wins only by exploiting non-uniform structure in the task distribution~\cite{wolpert1997nofreelunch}. The shortest adequate description of an arbitrary target is also not generally computable in the Kolmogorov sense~\cite{liVitanyi2008}. A foundry can dominate useful regions of software space, but it cannot own the shortest proof-carrying description of every possible future program.

This negative result is the boundary that makes the positive theory scientific. The claim is not that a model learns all possible future programs. The claim is that economically important software is concentrated in repeated behavior families, organizational patterns, integration rituals, schemas, policies, migrations, operational failures, and product conventions. That concentration is exactly what canonical code exploits~\cite{hindleNaturalness,dejavu2017,jupyterClones2020,clementsProductLines,dreamcoder}.

\subsection{The Horizon Equation}

Normalize today's raw all-in cost mass to one. Decompose it into an irreducible floor $\eta$ and disjoint reducible strata $q_1,\ldots,q_J$:
\begin{equation}
\eta+\sum_{j=1}^{J}q_j=1,
\qquad
q_j\ge 0,
\qquad
\eta\ge 0.
\label{eq:nah-cost-partition}
\end{equation}
Each $q_j$ is a removable source of accidental work: representation search, repository discovery, duplicate truth, proof-lane discovery, invalid edit attempts, routine behavior implementation, generated surface maintenance, retry loops, local architecture reconstruction, dependency rituals, migration folklore, or repeated review reasoning. Let $r_j\ge 1$ be the asymptotic reduction achieved on stratum $j$ by behavior cells, semantic patch cells, proof-carrying changes, canonical profiles, negative memory, renderers, and reusable evidence.

The residual cost fraction is
\begin{equation}
s_{\mathrm{NA}}=\eta+\sum_{j=1}^{J}\frac{q_j}{r_j},
\qquad
M_{\mathrm{NA}}=\frac{1}{s_{\mathrm{NA}}}.
\label{eq:nah-horizon-equation}
\end{equation}
Equation~\ref{eq:nah-horizon-equation} is the Amdahl law of inverse software~\cite{amdahl1967validity}. If two techniques reduce the same failed-search loop, they do not multiply; they compete for the same $q_j$. The exhausted-options limit is
\begin{equation}
\lim_{r_1,\ldots,r_J\rightarrow\infty}M_{\mathrm{NA}}=\frac{1}{\eta}.
\label{eq:nah-final-limit}
\end{equation}
The foundry does not escape the denominator. It drives every reducible term toward zero until only $\eta$ remains.

\subsection{Future-Adaptive Minimum Functional Description Length}

The irreducible floor is not merely source length. It is the cost of new information, acceptance, evidence, governance, and adaptability. An oracle-relative form is
\begin{align}
\mathrm{E}[C_F(Y)]\ge
\mathrm{E}\big[&\alpha K_{\oracle}(Y\mid M_t)
+\beta L_{\oracle}(Y)
+\gamma E_{\tau}(Y) \notag\\
&+\delta G(Y)+\zeta V_H(Y)\big].
\label{eq:nah-info-floor}
\end{align}
Here $K_{\oracle}(Y\mid M_t)$ is the target behavior not already implied by canonical memory, $L_{\oracle}(Y)$ is the description length of the acceptance oracle, $E_{\tau}(Y)$ is the minimum evidence burden, $G(Y)$ is governance and provenance burden, and $V_H(Y)$ is the option-value burden across horizon $H$. This is the \mfdl{} boundary in future-adaptive form: canonicalization can shorten descriptions and make proofs reusable, but it cannot make required new information disappear~\cite{grunwaldMDL,liVitanyi2008}.

\begin{figure*}[t]
  \centering
  \includegraphics[width=\textwidth]{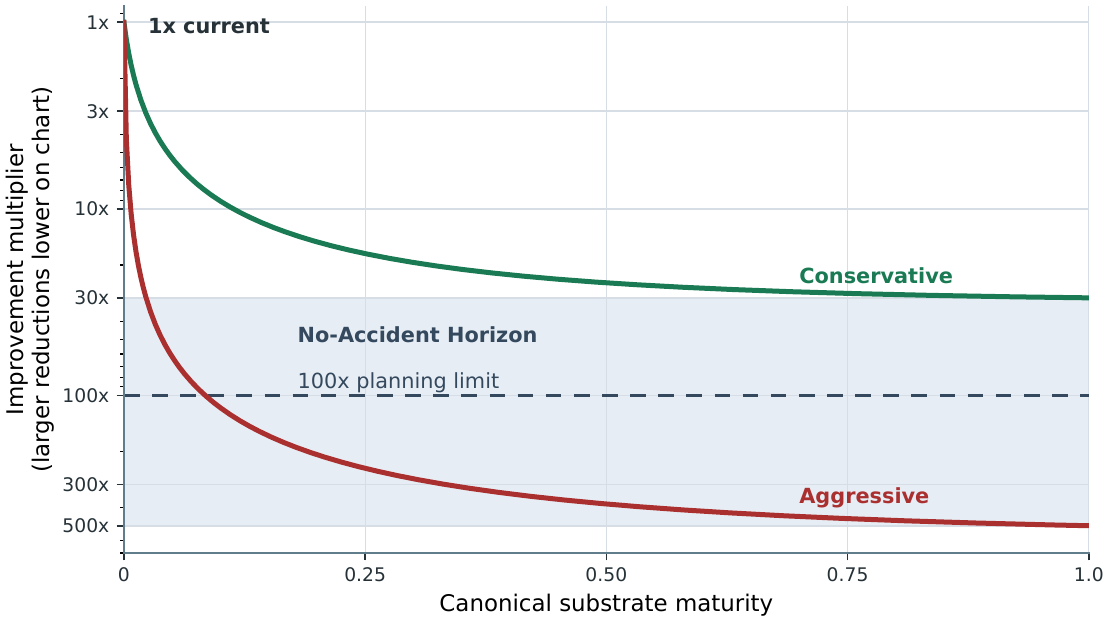}
  \caption{Convergence from current all-in verified-change cost toward the No-Accident Horizon. The conservative curve approaches a $30\times$ mature-foundry limit, the aggressive curve approaches a $500\times$ upper-tail limit for supported routine-product work, and the dashed line marks the $100\times$ theoretical planning horizon. The band is a theoretical exhausted-options estimate, not a measured result and not a distribution-free guarantee.}
  \label{fig:no-accident-horizon}
\end{figure*}

\subsection{Numerical Limit Estimate}

The strongest defensible numerical answer is a regime table, not a single slogan. Table~\ref{tab:no-accident-horizon-estimates} states the removable-work fraction $\Lambda_{\mathrm{NA}}$ and reciprocal multiplier $M_{\mathrm{NA}}$ for increasingly favorable assumptions.

\begin{table*}[t]
\caption{No-Accident Horizon: Final-Limit Estimates by Scope}
\label{tab:no-accident-horizon-estimates}
\centering
\footnotesize
\renewcommand{\arraystretch}{1.16}
\begin{tabularx}{\textwidth}{p{0.20\textwidth}X c c c}
\toprule
\textbf{Scope} & \textbf{Interpretation} & \textbf{Conservative} & \textbf{Central exhausted-options limit} & \textbf{Aggressive upper tail} \\
\midrule
Arbitrary programs as mathematical objects & Distribution-free future software, including adversarial and incompressible targets. & $0\%$ / $1\times$ & $0\%$ / $1\times$ & No positive universal bound \\
Broad economically requested software & Product apps plus systems, embedded, data, scientific, security, mobile, infrastructure, creative tools, and genuinely new algorithms. & $75\%$ / $4\times$ & $88\%$ / $8.3\times$ & $96\%$ / $25\times$ \\
Routine product and application software & SaaS, internal tools, resource lifecycles, auth, policy, billing, uploads, jobs, webhooks, audit, migrations, observability, deployment, and integrations. & $96.7\%$ / $30\times$ & $99.0\%$ / $100\times$ & $99.8\%$ / $500\times$ \\
Closed behavior-cell lanes & Narrow lanes where the future change is almost entirely cell selection, parameterization, renderer output, and reusable proof receipts. & $99.0\%$ / $100\times$ & $99.9\%$ / $1{,}000\times$ & $99.98\%$ / $5{,}000\times$ \\
\bottomrule
\end{tabularx}
\end{table*}

The table says three things. First, the universal mathematical problem has no positive guaranteed compression. Second, broad commercial software is compressible but still contains enough hardware, scientific, regulatory, adversarial, organizational, and algorithmic novelty that a universal $100\times$ all-in claim is not defensible. Third, the paper's real frontier is supported routine product software after behavior-genome maturity. In that regime, the strongest defensible final-limit planning number is
\begin{equation}
\Lambda_{\mathrm{NA,product}}^{\star}\approx 0.990
\qquad\Longleftrightarrow\qquad
M_{\mathrm{NA,product}}^{\star}\approx 100\times .
\label{eq:nah-product-star}
\end{equation}
A cautious theoretical band is
\begin{align}
\Lambda_{\mathrm{NA,product}}^{5\%\text{--}95\%}
&\in[0.967,0.998], \notag\\
M_{\mathrm{NA,product}}^{5\%\text{--}95\%}
&\in[30\times,500\times].
\label{eq:nah-product-band}
\end{align}
This is a theoretical limit estimate, not a measured result. It is stronger than the near-term central target because it assumes mature behavior cells, semantic patch cells, reusable proof receipts, negative memory, renderers, and broad amortization. It is weaker than an unqualified $1{,}000\times$ claim because Equation~\ref{eq:nah-final-limit} respects the residual floor.

\subsection{Falsification Rules}

The theory is designed to fail cleanly. Estimate $\eta$, $q_j$, and $r_j$ on paired raw/canonical repositories from the same lineage under the same model, issue distribution, hidden tests, proof lanes, reviewer rubric, and cost ledger. The No-Accident Horizon collapses into ordinary engineering efficiency if any of the following hold:
\begin{enumerate}
\item the irreducible floor $\eta$ remains above $3\%$ for routine product changes after mature canonicalization;
\item behavior-cell and semantic-patch coverage fail to exceed roughly $85\%$ of accepted routine change cost mass;
\item covered strata do not reach at least $100\times$ reduction in context, reasoning, invalid-action, retry, and proof-discovery cost;
\item proof, security, provenance, migration, or review overhead grows enough to erase the saved search cost;
\item canonical ports lose important compatibility ghosts or increase downstream defects at equal review standards;
\item foundry amortization does not compound across independent repositories.
\end{enumerate}
Conversely, the $100\times$ exhausted-options limit becomes conservative if $\eta<0.5\%$, covered routine-change mass exceeds $95\%$, cell-covered strata reach thousands-fold reductions, and proof receipts reuse across many unrelated product lineages without loss of behavior or provenance.

The final claim is therefore precise enough to defend and large enough to matter: the universe of economically important code can be built increasingly close to the No-Accident Horizon, but never beyond it. For arbitrary future programs, the guaranteed horizon is zero. For routine product/application software under a mature behavior-genome foundry, the strongest defensible final-limit estimate is about $100\times$ lower all-in cost per verified correct change, with a cautious $30\times$--$500\times$ theoretical band. Values above $1{,}000\times$ belong to closed behavior-cell lanes where the irreducible floor is below 0.1\%, not to unconstrained software as a whole.


%% file: sections/14_limits_and_conclusion.tex
\section{Limits and Conclusion}

This paper does not prove the full economic thesis. It does not prove behavior preservation at corpus scale, does not prove paired training-token reduction, does not prove that a 100B-class canonical specialist replaces all broad frontier coding systems, does not apply to all novel systems or unsupported domains, and does not prove that behavior cells cover 70\%--90\% of routine product behavior. The QLoRA pilot is deliberately narrower: canonical translated trajectories are learnable under a parameter-efficient adaptation setup, and the measured forbidden-language markers remain at zero. The rest of the thesis must be earned through the claim ledger in Table~\ref{tab:claim-status}, beginning with the same-model raw/canonical substrate test.

The research agenda is therefore concrete. Build paired raw/canonical repositories from the same lineage. Assign every source artifact a legal/provenance disposition. Port behavior only within declared contracts. Report preservation tiers, weak-oracle failures, compatibility ghosts, and accepted incompatibilities. Run the same broad model on raw and canonical tasks before training a specialist. Measure files opened, context tokens, reasoning tokens, tool calls, invalid edits, failed proof lanes, reviewer comments, wall-clock, and cost per accepted proof-carrying change. Then run paired scaling curves before claiming training-token compression.

The thesis remains bold because the target is not cleaner code. The target is a canonical mirror of the software universe: behavior preserved or dispositioned, routine product work collapsed into certified cells, changes expressed as typed operations, proof lanes generated and checked, provenance carried through every transformation, and runtime failures compiled into negative memory. Human code contains the behavior, edge cases, incidents, product judgment, and hard-won lessons that matter. Raw human representation is the wrong thing to imitate when it encodes local accident rather than durable behavior.

The theoretical limit is the No-Accident Horizon developed in Section~\ref{sec:no-accident-horizon}. In that limit, agents do not invent local architecture, rediscover repository folklore, guess migration law, or re-learn the same auth, billing, upload, policy, and audit patterns in every codebase. They operate a constrained software machine and spend compute only where novelty, evidence, governance, risk, and future optionality remain.

The end state is no accidental software: no accidental architecture, dependency choice, CI ritual, migration path, security policy, review burden, or repeated reasoning once the substrate has learned the law. This paper is a falsifiable program for replacing accidental software representation with a canonical, proof-carrying substrate for verified correct change. In the strongest form, every remaining unit of engineering buys novelty, judgment, evidence, governance, or safety.


%% file: compression_catalog.tex

\section{Supplemental Compression Catalog: Code-Space Gain Priors}
\label{sec:compression-catalog}

This supplemental catalog lists the major coding-work gains available once agent-first canonical code
is treated as a substrate rather than a style guide.  Each domain states
what is rebuilt, what becomes generated, what agent action paths
disappear, and conservative, central, and aggressive gain estimates.  The
gains \textbf{do not multiply cleanly}---many overlap, and several
domains share the same underlying token mass.  The composite effect must
be measured on paired raw/canonical corpora; the estimates presented
here are \emph{prior ranges and target hypotheses, not measured results}.
Where a per-domain source range such as $3\times$--$8\times$ is stated,
it denotes the ratio of raw-corpus tokens to canonical-corpus tokens
required to represent equivalent behavior.  Composite ranges use the
explicit denominator named in the row: representation space, action
space, training tokens, reasoning/tool/retry tokens, or cost per verified
correct change.


\subsection{Infrastructure Compression}

Infrastructure compression eliminates the representational cost of
decisions that carry no behavioral consequence: which language encodes a
REST handler, which framework provides routing, which folder tree
organizes source files.  These decisions dominate open-source corpora yet
contribute nothing to the space of behaviors a model must learn to
produce.  Domains~1--6 target this stratum.

\subsubsection{Language Role Collapse}
\label{ssec:lang-collapse}

\smallskip\noindent
\textbf{Approach.}\quad Rebuild service, UI, migration, and scripting roles into fixed profile lanes; generate adapters at profile boundaries; remove language-choice, runtime-choice, and cross-language repair paths.

The same backend behavior---HTTP routing, database access, queue
consumption, serialization---appears in Python/FastAPI, Java/Spring,
Node/Express, Ruby/Rails, PHP/Laravel, Go/Gin, and C\#/ASP.NET, among
others.  A canonical standard collapses these into one service/core lane
and one typed product-surface lane.  The model no longer spends capacity
learning seven syntactically distinct encodings of identical semantics.

\smallskip\noindent
\textbf{Estimates.}\quad
\conrange{\textbf{Conservative:} $4\times$--$6\times$}.\quad
Central: $6\times$--$12\times$.\quad
\aggrange{\textbf{Aggressive:} $12\times$--$20\times$}.

\subsubsection{Framework Collapse}
\label{ssec:framework-collapse}

\smallskip\noindent
\textbf{Approach.}\quad Rebuild routing, middleware, state, and persistence into one service/UI grammar; generate framework glue; remove local framework DSL choices and handler-shape variants.

Within a single language, framework conventions---middleware chains,
routing DSLs, ORM query builders, template engines, configuration
idioms---introduce a second layer of representational divergence.
Express, Django, Rails, Spring, and Laravel each impose a distinct
grammar over the same behavioral primitives.  Canonical conversion
replaces all framework-specific grammars with one canonical service
grammar, eliminating the combinatorial surface the model must memorize.

\smallskip\noindent
\textbf{Estimates.}\quad
\conrange{\textbf{Conservative:} $3\times$--$5\times$}.\quad
Central: $5\times$--$10\times$.\quad
\aggrange{\textbf{Aggressive:} $10\times$--$15\times$}.

\subsubsection{Generated Truth Collapse}
\label{ssec:generated-truth}

\smallskip\noindent
\textbf{Approach.}\quad Rebuild schemas as the source of truth; generate data transfer objects, clients, validators, fixtures, mocks, and docs; remove hand-edited projections and drift repairs.

Hand-written data transfer objects, API client stubs, request/response validators,
serialization fixtures, and test factories are all downstream projections
of a single schema contract.  In raw corpora these projections diverge,
drift, and duplicate.  Canonical form replaces them with contract-first
generation: one source of truth (an API schema or cell contract) produces
all projections deterministically.  The model learns to emit the contract,
not its many manual echoes.

\smallskip\noindent
\textbf{Estimates.}\quad
\conrange{\textbf{Conservative:} $2\times$--$3\times$}.\quad
Central: $3\times$--$6\times$.\quad
\aggrange{\textbf{Aggressive:} $6\times$--$10\times$}.

\subsubsection{Layout and Naming Collapse}
\label{ssec:layout-naming}

\smallskip\noindent
\textbf{Approach.}\quad Rebuild repositories into canonical paths and controlled vocabulary; generate owner maps and generated-zone manifests; remove file-hunt, synonym, and helper-placement choices.

Arbitrary folder structures (\texttt{src/controllers/} vs.\
\texttt{app/handlers/} vs.\ \texttt{lib/api/}), file-naming conventions
(\texttt{userService.ts} vs.\ \texttt{user\_service.py} vs.\
\texttt{UserService.java}), and organizational myths consume
representational capacity without encoding behavior.  A canonical
repository grammar fixes vocabulary and topology, converting layout
noise into a stable, predictable structure the model can exploit as
prior knowledge rather than re-learn per repository.

\smallskip\noindent
\textbf{Estimates.}\quad
\conrange{\textbf{Conservative:} $2\times$--$3\times$}.\quad
Central: $3\times$--$5\times$.\quad
\aggrange{\textbf{Aggressive:} $5\times$--$8\times$}.

\subsubsection{Dependency Collapse}
\label{ssec:dependency-collapse}

\smallskip\noindent
\textbf{Approach.}\quad Rebuild integrations behind governed adapters; generate configuration, retry, error, and observability wrappers; remove bespoke vendor glue and unsafe upgrade routes.

Every production codebase wraps third-party services---Stripe, S3,
GitHub, Slack, OpenAI, Redis, email providers, observability
backends---in locally invented adapter layers with inconsistent error
handling, retry policies, and configuration surfaces.  Canonical adapters
provide governed wrappers with standard interfaces.  The model learns one
adapter contract per external service rather than hundreds of bespoke
wrappers, and local glue code vanishes entirely.

\smallskip\noindent
\textbf{Estimates.}\quad
\conrange{\textbf{Conservative:} $2\times$--$3\times$}.\quad
Central: $3\times$--$5\times$.\quad
\aggrange{\textbf{Aggressive:} $5\times$--$8\times$}.

\subsubsection{Build and Test Collapse}
\label{ssec:build-test}

\smallskip\noindent
\textbf{Approach.}\quad Rebuild local and CI validation as proof lanes; generate GitHub/GitLab workflow settings, artifacts, and required checks; remove build folklore and permission guesswork.

Custom shell scripts, undocumented CI gate configurations, Makefile
folklore, local environment assumptions, and ad-hoc test harnesses
constitute a significant fraction of repository tokens.  Canonical form
replaces them with standard proof lanes: a fixed set of commands that
produce expected artifacts (type-checked binary, migration proof, test
evidence, security scan).  The model learns the proof protocol, not the
archaeology of each project's build system.

\smallskip\noindent
\textbf{Estimates.}\quad
\conrange{\textbf{Conservative:} $2\times$--$3\times$}.\quad
Central: $2\times$--$4\times$.\quad
\aggrange{\textbf{Aggressive:} $4\times$--$6\times$}.

\bigskip
\begin{table*}[t]
\centering
\caption{Infrastructure Compression Domains (1--6): Gain Estimates}
\label{tab:infra-compression}
{\footnotesize
\begin{tabularx}{\textwidth}{@{} l X >{\color{nhConservative}}c c >{\color{nhAggressive}}c @{}}
\toprule
\textbf{\#} & \textbf{Domain} & \conhead & \textbf{Central} & \agghead \\
\midrule
1 & Language Role Collapse        & $4\times$--$6\times$   & $6\times$--$12\times$  & $12\times$--$20\times$ \\
2 & Framework Collapse            & $3\times$--$5\times$   & $5\times$--$10\times$  & $10\times$--$15\times$ \\
3 & Generated Truth Collapse      & $2\times$--$3\times$   & $3\times$--$6\times$   & $6\times$--$10\times$  \\
4 & Layout/Naming Collapse        & $2\times$--$3\times$   & $3\times$--$5\times$   & $5\times$--$8\times$   \\
5 & Dependency Collapse           & $2\times$--$3\times$   & $3\times$--$5\times$   & $5\times$--$8\times$   \\
6 & Build/Test Collapse           & $2\times$--$3\times$   & $2\times$--$4\times$   & $4\times$--$6\times$   \\
\bottomrule
\end{tabularx}
}
\end{table*}


\subsection{Behavior Cell Compression}

Behavior cells are the canonical unit of application logic: a named,
typed, proof-carrying declaration of a behavioral intent that the
canonical runtime can instantiate, compose, and verify.  Where
infrastructure compression removes \emph{representational} divergence,
cell compression removes \emph{behavioral} divergence---the phenomenon
whereby identical application semantics (pagination, authentication,
billing) are re-implemented from scratch in every codebase with
local variation that encodes no new information.  Domains~7--20 target
this stratum.  To keep the catalog compact, several high-win cells are
folded into adjacent domains: tenant/org/user belongs to identity and
policy, feature flags and experiments belong to configuration policy,
cache/invalidation belongs to governed dependency and observability
contracts, and UI resource state belongs to list/search cells. Patch
intermediate representation, reasoning digests, supply-chain closure,
and notebook-to-pipeline compression are treated separately in the
meta-compression layer because their denominators are action,
reasoning, governance, and workflow entropy rather than source tokens
alone.

\subsubsection{Resource Lifecycle Cells}
\label{ssec:crud-cells}

\smallskip\noindent
\textbf{Approach.}\quad Rebuild resource lifecycle work as typed cell declarations; generate routes, SQL, UI state, tests, and fixtures; remove handwritten create/read/update/delete surfaces and repeated controller edits.

Create, read, update, delete, list, archive, restore, and status
transition constitute the single most repeated pattern in product
software.  In raw corpora, each resource lifecycle surface is hand-written with
per-project validation, authorization, serialization, and error handling.
Canonical resource lifecycle cells reduce this to a typed resource declaration: the
model emits cell parameters (fields, transitions, access policies)
rather than writing route handlers, SQL queries, and test fixtures.

\smallskip\noindent
\textbf{Estimates.}\quad
\conrange{\textbf{Conservative:} $3\times$--$8\times$}.\quad
Central: $8\times$--$20\times$.\quad
\aggrange{\textbf{Aggressive:} $20\times$--$50\times$}.

\subsubsection{Authentication / Identity / Policy Cells}
\label{ssec:auth-cells}

\smallskip\noindent
\textbf{Approach.}\quad Rebuild identity, tenant, organization, session, and policy behavior as governed cells; generate guards, recovery flows, fixtures, and negative tests; remove bespoke auth plumbing.

Login flows, session management, token issuance and validation, external
identity-provider integration, role-based access control,
attribute-based access control, permission checks, password reset,
multi-factor authentication, and account recovery are repeated in nearly
every production application---with dangerous variation.  Security-critical
code is the worst candidate for bespoke re-implementation and the best
candidate for cell compression.  A canonical identity cell encapsulates
the full policy surface; the model configures it rather than writing
cryptographic plumbing.

\smallskip\noindent
\textbf{Estimates.}\quad
\conrange{\textbf{Conservative:} $3\times$--$10\times$}.\quad
Central: $10\times$--$30\times$.\quad
\aggrange{\textbf{Aggressive:} $30\times$--$100\times$}.

\subsubsection{Form and Input Validation Cells}
\label{ssec:validation-cells}

\smallskip\noindent
\textbf{Approach.}\quad Rebuild validation as schema-owned constraints; generate frontend, API, database, and test projections; remove four-layer drift and hand-maintained form logic.

Validation logic is duplicated across at least four layers in typical
applications: frontend form validators, API request schemas, database
constraints, and test assertions.  Each layer uses a different validation
DSL, and drift between layers is a persistent source of bugs.  Canonical
validation cells generate all four projections from a single schema,
eliminating cross-layer inconsistency and the representational cost of
maintaining parallel validation grammars.

\smallskip\noindent
\textbf{Estimates.}\quad
\conrange{\textbf{Conservative:} $2\times$--$5\times$}.\quad
Central: $5\times$--$15\times$.\quad
\aggrange{\textbf{Aggressive:} $15\times$--$30\times$}.

\subsubsection{Pagination / Search / Filter Cells}
\label{ssec:pagination-cells}

\smallskip\noindent
\textbf{Approach.}\quad Rebuild list/search behavior as a resource-state cell; generate durable-data queries, API cursors, product-surface state, empty/error/loading views, and tests; remove custom list implementations.

Paginated lists with sorting, filtering, full-text search, cursor-based
or offset pagination, and search index synchronization represent massive
UI/API/database repetition.  The same behavioral contract---``return a
filtered, sorted, paginated view of a resource collection''---is
implemented independently at every layer with per-project conventions.
A canonical pagination cell parameterizes the contract once; all layers
are generated.

\smallskip\noindent
\textbf{Estimates.}\quad
\conrange{\textbf{Conservative:} $3\times$--$8\times$}.\quad
Central: $8\times$--$20\times$.\quad
\aggrange{\textbf{Aggressive:} $20\times$--$50\times$}.

\subsubsection{SQL Migration and Data Invariant Cells}
\label{ssec:migration-cells}

\smallskip\noindent
\textbf{Approach.}\quad Rebuild data change as declared invariants, lock budgets, and expand/contract plans; generate SQL and replay checks; remove opaque migration scripts and unsafe rollbacks.

Expand/contract migrations, data backfills, rollback plans, data-shape
invariants, and lock-budget management carry high failure cost and high
repetition.  In raw corpora, migration files are opaque SQL scripts with
no machine-readable invariant or proof of safety.  Canonical migration
cells declare the schema transition, invariant constraints, and
acceptable lock budgets; the runtime generates the SQL and verifies
safety properties before execution.

\smallskip\noindent
\textbf{Estimates.}\quad
\conrange{\textbf{Conservative:} $2\times$--$5\times$}.\quad
Central: $5\times$--$10\times$.\quad
\aggrange{\textbf{Aggressive:} $10\times$--$20\times$}.

\subsubsection{Background Job / Queue / Scheduler Cells}
\label{ssec:job-cells}

\smallskip\noindent
\textbf{Approach.}\quad Rebuild async work as idempotent job contracts; generate retries, dead-letter routing, schedules, metrics, and tests; remove queue-specific reliability rewrites.

Retry policies, idempotency guarantees, exponential backoff, dead-letter
queue routing, scheduled task management, and concurrency limits
constitute the reliability layer of asynchronous processing.  The same
reliability logic is re-implemented in every codebase that uses
background jobs, with subtle bugs in retry semantics and idempotency
checks.  Canonical job cells declare the reliability contract; the
runtime enforces it.

\smallskip\noindent
\textbf{Estimates.}\quad
\conrange{\textbf{Conservative:} $2\times$--$5\times$}.\quad
Central: $5\times$--$15\times$.\quad
\aggrange{\textbf{Aggressive:} $15\times$--$30\times$}.

\subsubsection{File Upload and Media Cells}
\label{ssec:media-cells}

\smallskip\noindent
\textbf{Approach.}\quad Rebuild uploads as storage and policy declarations; generate signed URLs, scanners, transforms, fixtures, and audit events; remove ad hoc storage/security code.

Signed URL generation, virus scanning, image processing pipelines,
storage backend abstraction, content-type validation, and access control
for uploaded media combine security, storage, and processing complexity
that is repeated in nearly every user-facing application.  Canonical
media cells declare upload constraints, processing steps, and access
policies; infrastructure details are resolved by the runtime.

\smallskip\noindent
\textbf{Estimates.}\quad
\conrange{\textbf{Conservative:} $3\times$--$8\times$}.\quad
Central: $8\times$--$20\times$.\quad
\aggrange{\textbf{Aggressive:} $20\times$--$40\times$}.

\subsubsection{Billing / Subscription / Payment Cells}
\label{ssec:billing-cells}

\smallskip\noindent
\textbf{Approach.}\quad Rebuild payments as subscription and ledger state machines; generate webhook handlers, idempotency, reconciliation, fixtures, and alerts; remove brittle payment glue.

Subscription lifecycle management, invoice preview and proration,
payment method lifecycle, webhook handling for Stripe and other payment
providers, dunning flows, and tax calculation represent a common SaaS
surface with dangerous edge cases in currency handling, idempotency, and
state reconciliation.  Canonical billing cells parameterize the
subscription model and payment provider; the full webhook surface,
idempotency layer, and state machine are generated.

\smallskip\noindent
\textbf{Estimates.}\quad
\conrange{\textbf{Conservative:} $3\times$--$10\times$}.\quad
Central: $10\times$--$30\times$.\quad
\aggrange{\textbf{Aggressive:} $30\times$--$100\times$}.

\subsubsection{Notification / Email / Template Cells}
\label{ssec:notification-cells}

\smallskip\noindent
\textbf{Approach.}\quad Rebuild outbound messaging as channel and template policy; generate provider adapters, preferences, retries, and observability; remove local email/SMS/push wrappers.

Multi-channel notification dispatch (email, SMS, push, in-app), delivery
tracking, template rendering with variable substitution, user preference
management, and retry logic for transient delivery failures.  Canonical
notification cells declare channels, templates, and delivery policies;
the runtime handles provider integration and observability.

\smallskip\noindent
\textbf{Estimates.}\quad
\conrange{\textbf{Conservative:} $2\times$--$5\times$}.\quad
Central: $5\times$--$15\times$.\quad
\aggrange{\textbf{Aggressive:} $15\times$--$30\times$}.

\subsubsection{Webhook and Integration Cells}
\label{ssec:webhook-cells}

\smallskip\noindent
\textbf{Approach.}\quad Rebuild integrations as signed event contracts; generate verification, replay, deduplication, dispatch, and fixtures; remove provider-specific webhook boilerplate.

Inbound webhook verification, signature validation, idempotent event
processing, replay handling, deduplication, and provider-specific payload
parsing.  Integration-heavy applications repeat this pattern per external
provider, each with a different signing scheme and payload format.
Canonical integration cells declare the provider contract and
verification method; the idempotency and processing pipeline are
standard.

\smallskip\noindent
\textbf{Estimates.}\quad
\conrange{\textbf{Conservative:} $2\times$--$5\times$}.\quad
Central: $5\times$--$15\times$.\quad
\aggrange{\textbf{Aggressive:} $15\times$--$30\times$}.

\subsubsection{Audit Log and Event History Cells}
\label{ssec:audit-cells}

\smallskip\noindent
\textbf{Approach.}\quad Rebuild audit as append-only event policy; generate event writers, retention, queries, and tamper checks; remove scattered compliance logging.

Append-only event streams, compliance audit trails,
actor/action/resource/timestamp recording, and tamper-evident log
integrity are required in compliance-heavy systems (healthcare, finance,
government).  In raw corpora, audit logging is either absent, ad-hoc, or
inconsistently applied.  Canonical audit cells declare the event schema
and retention policy; the runtime guarantees append-only semantics and
query access.

\smallskip\noindent
\textbf{Estimates.}\quad
\conrange{\textbf{Conservative:} $2\times$--$5\times$}.\quad
Central: $5\times$--$10\times$.\quad
\aggrange{\textbf{Aggressive:} $10\times$--$20\times$}.

\subsubsection{Observability / Logging / Tracing Cells}
\label{ssec:observability-cells}

\smallskip\noindent
\textbf{Approach.}\quad Rebuild telemetry as a cross-cutting contract; generate logs, spans, metrics, health checks, and correlation IDs; remove inconsistent instrumentation edits.

Structured logging, distributed tracing with context propagation,
metrics collection and export, request-ID propagation across service
boundaries, and health-check endpoints.  These cross-cutting concerns
are woven inconsistently through raw codebases, often with conflicting
log formats and missing trace context.  Canonical observability cells
declare the instrumentation contract; the runtime injects it uniformly.

\smallskip\noindent
\textbf{Estimates.}\quad
\conrange{\textbf{Conservative:} $2\times$--$3\times$}.\quad
Central: $3\times$--$8\times$.\quad
\aggrange{\textbf{Aggressive:} $8\times$--$15\times$}.

\subsubsection{Configuration / Secrets / Environment Cells}
\label{ssec:config-cells}

\smallskip\noindent
\textbf{Approach.}\quad Rebuild configuration, secrets, feature flags, and experiments as typed policy; generate env validation, rotation checks, flag gates, and rollout tests; remove .env folklore.

Secret rotation, environment-specific configuration, feature flag
management, configuration validation at startup, and secret-leakage
prevention.  In raw corpora, \texttt{.env} files, hardcoded secrets,
and undocumented environment variables constitute a persistent security
and reliability risk.  Canonical configuration cells declare the
configuration schema with types, defaults, and secret classifications;
the runtime validates at boot and enforces rotation policies.

\smallskip\noindent
\textbf{Estimates.}\quad
\conrange{\textbf{Conservative:} $2\times$--$4\times$}.\quad
Central: $4\times$--$8\times$.\quad
\aggrange{\textbf{Aggressive:} $8\times$--$15\times$}.

\subsubsection{Error Handling and Result Envelope Cells}
\label{ssec:error-cells}

\smallskip\noindent
\textbf{Approach.}\quad Rebuild failure handling as typed result envelopes; generate error codes, retry classifications, UI messages, and tests; remove stringly exceptions and null-path ambiguity.

Typed error envelopes, machine-readable error codes, retry
classification (transient vs.\ permanent), user-facing vs.\ internal
error separation, and structured error metadata.  Raw codebases exhibit
a chaotic mixture of exceptions, HTTP status codes, null returns, and
string error messages.  Canonical error cells impose a typed result
algebra; the model learns to classify and route errors rather than
invent ad-hoc handling per call site.

\smallskip\noindent
\textbf{Estimates.}\quad
\conrange{\textbf{Conservative:} $2\times$--$4\times$}.\quad
Central: $4\times$--$8\times$.\quad
\aggrange{\textbf{Aggressive:} $8\times$--$12\times$}.

\bigskip
\begin{table*}[t]
\centering
\caption{Behavior Cell Compression Domains (7--20): Gain Estimates}
\label{tab:cell-compression}
{\footnotesize
\begin{tabularx}{\textwidth}{@{} l X >{\color{nhConservative}}c c >{\color{nhAggressive}}c @{}}
\toprule
\textbf{\#} & \textbf{Domain} & \conhead & \textbf{Central} & \agghead \\
\midrule
7  & Resource Lifecycle                     & $3\times$--$8\times$    & $8\times$--$20\times$    & $20\times$--$50\times$    \\
8  & Authentication / Identity / Policy     & $3\times$--$10\times$   & $10\times$--$30\times$   & $30\times$--$100\times$   \\
9  & Form / Input Validation                & $2\times$--$5\times$    & $5\times$--$15\times$    & $15\times$--$30\times$    \\
10 & Pagination / Search / Filter           & $3\times$--$8\times$    & $8\times$--$20\times$    & $20\times$--$50\times$    \\
11 & SQL Migration / Data Invariants        & $2\times$--$5\times$    & $5\times$--$10\times$    & $10\times$--$20\times$    \\
12 & Background Job / Queue / Scheduler     & $2\times$--$5\times$    & $5\times$--$15\times$    & $15\times$--$30\times$    \\
13 & File Upload / Media                    & $3\times$--$8\times$    & $8\times$--$20\times$    & $20\times$--$40\times$    \\
14 & Billing / Subscription / Payment       & $3\times$--$10\times$   & $10\times$--$30\times$   & $30\times$--$100\times$   \\
15 & Notification / Email / Template        & $2\times$--$5\times$    & $5\times$--$15\times$    & $15\times$--$30\times$    \\
16 & Webhook / Integration                  & $2\times$--$5\times$    & $5\times$--$15\times$    & $15\times$--$30\times$    \\
17 & Audit Log / Event History              & $2\times$--$5\times$    & $5\times$--$10\times$    & $10\times$--$20\times$    \\
18 & Observability / Logging / Tracing      & $2\times$--$3\times$    & $3\times$--$8\times$     & $8\times$--$15\times$     \\
19 & Config / Secrets / Environment         & $2\times$--$4\times$    & $4\times$--$8\times$     & $8\times$--$15\times$     \\
20 & Error Handling / Result Envelope       & $2\times$--$4\times$    & $4\times$--$8\times$     & $8\times$--$12\times$     \\
\bottomrule
\end{tabularx}
}
\end{table*}


\subsection{Meta-Compression}

Meta-compression operates not on source tokens but on the
\emph{action space and reasoning space} of the agent itself.  Where infrastructure and cell
compression shrink the representation a model must learn to read and
write, meta-compression shrinks the space of \emph{trajectories} a model
must explore during generation.  The gains here are measured not in
token-count ratios but in the reduction of invalid action paths, wasted
inference steps, unproductive exploration, review burden, supply-chain
ambiguity, and workflow reinvention.  Domains~21--30 target
this stratum.

\subsubsection{Repair-Path Collapse}
\label{ssec:repair-path}

\smallskip\noindent
\textbf{Approach.}\quad Rebuild patching as constrained edit policy; generate legal edit maps, proof routes, and no-edit zones; remove arbitrary file, command, migration, and dependency choices.

In a canonical environment, the agent no longer selects among arbitrary
files to edit, frameworks to invoke, proof commands to run, generated
artifacts to modify, or migration strategies to attempt.  The canonical
grammar constrains every dimension of the action space simultaneously:
which files are mutable, which cells accept parameters, which proof
lanes must pass, which artifacts are generated.  Invalid edit
paths---the overwhelming majority of the raw action space---vanish
entirely.  This is the single largest compression domain by
estimated magnitude, because the space of \emph{wrong} actions in
unconstrained codebases is combinatorially vast.

\smallskip\noindent
\textbf{Estimates.}\quad
\conrange{\textbf{Conservative:} $10\times$--$100\times$}.\quad
Central: $100\times$--$10{,}000\times$.\quad
\aggrange{\textbf{Aggressive:} $10{,}000\times$--$100{,}000\times$}.

\subsubsection{Negative-Path Curriculum}
\label{ssec:negative-path}

\smallskip\noindent
\textbf{Approach.}\quad Rebuild mistakes as labeled curriculum; generate invalid edits, missing proofs, unsafe migrations, and generated-zone violations; remove ambiguous definitions of wrongness.

Canonical form enables a training curriculum that includes not only
correct canonical code but also \textbf{labeled negative examples}:
editing generated files, applying patches to the wrong architectural
layer, writing unsafe migrations without lock-budget declarations,
duplicating generated data-transfer-object definitions, and skipping
proof lanes.  These negatives are cheap to produce in a canonical
environment (any violation of the grammar is a negative) and expensive
to produce in raw corpora (where ``wrong'' is ambiguous).  The model
learns the \emph{boundary} of correct behavior, not just its interior.

\smallskip\noindent
\textbf{Estimates.}\quad
\conrange{\textbf{Conservative:} $2\times$--$5\times$}.\quad
Central: $5\times$--$15\times$.\quad
\aggrange{\textbf{Aggressive:} $15\times$--$50\times$}.

\subsubsection{Proof-Carrying Change Objects}
\label{ssec:proof-carrying}

\smallskip\noindent
\textbf{Approach.}\quad Rebuild diffs as structured change objects and patch IR; generate intent, affected cells, proofs, receipts, and rollback evidence; remove bare diff-only completion paths.

In canonical form, a patch is not a raw diff.  It is a structured change
object comprising: the declared intent, the affected cells, the generated
delta, the migration proof, the security proof, and the test evidence.
The model learns to produce the \emph{proof route}---the full
trajectory from intent to verified change---rather than a bare textual
patch whose correctness must be inferred post-hoc.  This transforms
code generation from a single-shot prediction problem into a structured
reasoning chain with verifiable intermediate steps.

\smallskip\noindent
\textbf{Estimates.}\quad
\conrange{\textbf{Conservative:} $2\times$--$5\times$}.\quad
Central: $5\times$--$10\times$.\quad
\aggrange{\textbf{Aggressive:} $10\times$--$20\times$}.

\subsubsection{Reasoning Digest and Plan-Cache Compression}
\label{ssec:reasoning-digests}

\smallskip\noindent
\textbf{Approach.}\quad Rebuild repeated agent cognition as versioned reasoning digests; generate task routes, known failure modes, proof obligations, and repair playbooks; remove repeated repo archaeology.

Current agents spend inference tokens rediscovering architecture,
ownership, generated zones, tests, migration rituals, and safe repair
strategies.  Canonical repositories should store a compact reasoning
map: task class to affected cells, issue type to legal edit set, failure
signature to repair strategy, policy delta to required test matrix, and
dependency update to compatibility checks.  This turns repeated thought
into substrate memory.

\smallskip\noindent
\textbf{Estimates.}\quad
\conrange{\textbf{Conservative:} $3\times$--$8\times$}.\quad
Central: $8\times$--$50\times$.\quad
\aggrange{\textbf{Aggressive:} $50\times$--$200\times$}.

\subsubsection{Semantic Patch Cells}
\label{ssec:semantic-patch-cells}

\smallskip\noindent
\textbf{Approach.}\quad Rebuild recurring changes as typed transformations; generate blast radius, preconditions, proofs, tests, and rollbacks; remove raw-diff invention for common edits.

Many changes recur across repositories: add field, migrate nullable to
required, add permission edge, rotate secret, add idempotency key,
upgrade dependency safely, split table, add audit trail, deprecate API
field, or convert sync work to an idempotent job.  Semantic patch cells
make those changes first-class.  The agent binds parameters and
discharges obligations instead of inventing edits file by file.

\smallskip\noindent
\textbf{Estimates.}\quad
\conrange{\textbf{Conservative:} $10\times$--$50\times$}.\quad
Central: $50\times$--$1{,}000\times$.\quad
\aggrange{\textbf{Aggressive:} $1{,}000\times$--$10{,}000\times$}.

\subsubsection{Constrained Edit Grammar}
\label{ssec:constrained-edit-grammar}

\smallskip\noindent
\textbf{Approach.}\quad Rebuild editing as legal typed operations over behavior IR, cells, schemas, and policies; reject illegal files, operations, and proof omissions before generation.

Repair-path collapse removes many wrong trajectories after the agent
chooses a path.  A constrained edit grammar removes them before choice.
Generated files are immutable; migrations must declare lock budgets and
rollback; policy changes must emit negative tests; dependency changes
must route through governed upgrade lanes.  The agent sees a small legal
action set instead of a whole filesystem.

\smallskip\noindent
\textbf{Estimates.}\quad
\conrange{\textbf{Conservative:} $100\times$--$1{,}000\times$}.\quad
Central: $1{,}000\times$--$100{,}000\times$.\quad
\aggrange{\textbf{Aggressive:} $100{,}000\times$+}.

\subsubsection{Proof-Lane Receipt Reuse}
\label{ssec:proof-receipts}

\smallskip\noindent
\textbf{Approach.}\quad Rebuild verification as reusable receipts; generate migration replay, rollback evidence, policy checks, and security receipts; remove repeated proof interpretation and review reconstruction.

Proof lanes define commands.  Receipt reuse defines the evidence object
that survives the command: what was checked, which obligations changed,
which logs matter, what failed before repair, and what rollback means.
Reviewers and agents inspect the receipt instead of reconstructing proof
meaning from raw logs.

\smallskip\noindent
\textbf{Estimates.}\quad
\conrange{\textbf{Conservative:} $3\times$--$8\times$}.\quad
Central: $8\times$--$30\times$.\quad
\aggrange{\textbf{Aggressive:} $30\times$--$100\times$}.

\subsubsection{Supply-Chain and Dependency Closure}
\label{ssec:supply-chain-closure}

\smallskip\noindent
\textbf{Approach.}\quad Rebuild package selection as governed capability resolution; generate adapters, provenance, license, bill-of-materials, vulnerability, upgrade, and rollback evidence; remove arbitrary package choice.

Dependencies are hidden code expansion.  A single import can carry
transitive code, build scripts, licenses, vulnerabilities, and runtime
behavior.  Canonical form should resolve verified capabilities rather
than package names: password hashing, email send, payment webhook,
rate limit, or object storage.  Each capability has approved providers,
adapters, provenance, proof lanes, and upgrade routes.

\smallskip\noindent
\textbf{Estimates.}\quad
\conrange{\textbf{Conservative:} $2\times$--$5\times$}.\quad
Central: $5\times$--$20\times$.\quad
\aggrange{\textbf{Aggressive:} $20\times$--$100\times$}.

\subsubsection{Notebook-to-Pipeline Compression}
\label{ssec:notebook-pipeline}

\smallskip\noindent
\textbf{Approach.}\quad Rebuild exploratory notebooks as typed data pipelines; generate environment locks, provenance, tests, parameter cells, and scheduled jobs; remove copy-pasted exploratory state.

Notebook-heavy software often stores executable history rather than
reproducible behavior.  Canonical conversion separates exploration from
pipeline truth: cells become typed data steps, parameters become schema,
plots and reports become generated projections, and environments become
locked artifacts.  The agent learns the pipeline graph rather than the
accidental order of notebook execution.

\smallskip\noindent
\textbf{Estimates.}\quad
\conrange{\textbf{Conservative:} $3\times$--$10\times$}.\quad
Central: $10\times$--$50\times$.\quad
\aggrange{\textbf{Aggressive:} $50\times$--$200\times$}.

\subsubsection{Runtime and Negative Memory Compression}
\label{ssec:runtime-negative-memory}

\smallskip\noindent
\textbf{Approach.}\quad Rebuild incidents, reverts, failed patches, hidden compatibility, and production traces as living invariants and forbidden plans; remove repeated rediscovery of bad worlds.

The raw corpus mostly preserves accepted code, but software expertise
also lives in what failed: unsafe migrations, auth bypasses, flaky
tests, reverted patches, support incidents, and production-only edge
cases.  Canonical memory turns those failures into anti-examples,
negative tests, proof obligations, and reasoning warnings.  Every
failure becomes a compression artifact.

\smallskip\noindent
\textbf{Estimates.}\quad
\conrange{\textbf{Conservative:} $2\times$--$5\times$}.\quad
Central: $5\times$--$25\times$.\quad
\aggrange{\textbf{Aggressive:} $25\times$--$100\times$}.

\bigskip
\begin{table*}[t]
\centering
\caption{Meta-Compression Domains (21--30): Gain Estimates}
\label{tab:meta-compression}
{\footnotesize
\begin{tabularx}{\textwidth}{@{} l X >{\color{nhConservative}}c c >{\color{nhAggressive}}c @{}}
\toprule
\textbf{\#} & \textbf{Domain} & \conhead & \textbf{Central} & \agghead \\
\midrule
21 & Repair-Path Collapse           & $10\times$--$100\times$       & $100\times$--$10{,}000\times$       & $10{,}000\times$--$100{,}000\times$ \\
22 & Negative-Path Curriculum       & $2\times$--$5\times$          & $5\times$--$15\times$               & $15\times$--$50\times$              \\
23 & Proof-Carrying Change Objects  & $2\times$--$5\times$          & $5\times$--$10\times$               & $10\times$--$20\times$              \\
24 & Reasoning Digests / Plan Cache & $3\times$--$8\times$          & $8\times$--$50\times$               & $50\times$--$200\times$             \\
25 & Semantic Patch Cells           & $10\times$--$50\times$        & $50\times$--$1{,}000\times$         & $1{,}000\times$--$10{,}000\times$   \\
26 & Constrained Edit Grammar       & $100\times$--$1{,}000\times$  & $1{,}000\times$--$100{,}000\times$  & $100{,}000\times$+                  \\
27 & Proof-Lane Receipt Reuse       & $3\times$--$8\times$          & $8\times$--$30\times$               & $30\times$--$100\times$             \\
28 & Supply-Chain / Dependency Closure & $2\times$--$5\times$       & $5\times$--$20\times$               & $20\times$--$100\times$             \\
29 & Notebook-to-Pipeline           & $3\times$--$10\times$         & $10\times$--$50\times$              & $50\times$--$200\times$             \\
30 & Runtime / Negative Memory      & $2\times$--$5\times$          & $5\times$--$25\times$               & $25\times$--$100\times$             \\
\bottomrule
\end{tabularx}
}
\end{table*}


\subsubsection{Composite Gain Estimates}
\label{ssec:composite-gains}

The per-domain estimates above cannot be multiplied na\"ively: language
collapse and framework collapse share token mass; resource lifecycle cells and
validation cells overlap on the same source files; repair-path collapse
subsumes portions of every other domain.  Table~\ref{tab:composite}
presents composite estimates across five measurement axes, derived from
conservative overlap-discounting of the individual domains.  These
composites represent the full-stack prior: the expected gain when
\emph{all} canonical transformations are applied simultaneously to a
representative product codebase.

\begin{table*}[t]
\centering
\caption{Composite Gain Estimates Across All 30 Compression Domains (Denominator-Specific; Not Multiplicative)}
\label{tab:composite}
{\footnotesize
\begin{tabularx}{\textwidth}{@{} X >{\color{nhConservative}}c c >{\color{nhAggressive}}c @{}}
\toprule
\textbf{Measurement Axis} & \conhead & \textbf{Central} & \agghead \\
\midrule
Literal source-token reduction
    & $2\times$--$8\times$
    & $3\times$--$10\times$
    & $10\times$--$30\times$ \\
Effective representation-space reduction
    & $20\times$--$40\times$
    & $40\times$--$150\times$
    & $100\times$--$300\times$ \\
Agent action-space reduction
    & $100\times$--$1{,}000\times$
    & $1{,}000\times$--$100{,}000\times$
    & $100{,}000\times$+ \\
Training-token efficiency
    & $10\times$--$30\times$
    & $30\times$--$150\times$
    & $150\times$--$1{,}000\times$ \\
Reasoning/tool/retry-token reduction
    & $3\times$--$10\times$
    & $10\times$--$100\times$
    & $100\times$--$10{,}000\times$ \\
Cost per verified correct change
    & $3\times$--$10\times$
    & $10\times$--$50\times$
    & $50\times$--$1{,}000\times$ \\
\bottomrule
\end{tabularx}
}
\end{table*}

The distinction between \emph{literal source-token reduction} and
\emph{effective representation-space reduction} is critical.  Source
tokens measure the physical size of the corpus; representation space
measures the number of \emph{distinct behavioral encodings} the model
must learn to produce.  A $5\times$ source-token reduction that also
collapses seven languages into one yields a far larger effective
reduction, because the model no longer maintains seven parallel
decoders for the same semantic space.  Agent action-space reduction is
larger still, because it compounds source compression with the
elimination of invalid edit trajectories (Domain~21).  Reasoning-token
reduction is a separate denominator: it measures how often the agent
must rediscover architecture, tests, proof routes, repair strategy,
dependency law, and review expectations that can instead be compiled
into reasoning digests, proof receipts, and semantic patch cells.


\subsubsection{Theoretical Limit: Minimum Functional Description Length}
\label{ssec:mfdl}

The compression frontier converges toward a theoretical limit we term
the \textbf{Minimum Functional Description Length}: the shortest
canonical specification, plus proofs, plus renderer needed to produce a
fully working system.  Minimum Functional Description Length is not a code-golf metric; it is the
information-theoretic minimum over the space of behavioral
specifications that can be mechanically verified and rendered into
executable artifacts.  For routine product software---the SaaS
platforms, internal tools, and resource-heavy services that constitute the
majority of commercial codebases---we estimate that 70--90\% of behavior
may eventually be expressible as cell composition, generated surfaces,
and policy configuration.  The remaining 10--30\% is true
domain-specific logic: novel algorithms, unique business rules, and
irreducible problem-specific reasoning that no canonical grammar can
absorb without becoming domain-specific itself.

The theoretical limit is not shorter code.  \textbf{The theoretical
limit is no bespoke code unless the behavior is novel.}